\documentclass[12pt,twoside]{article}

\usepackage{amssymb}
\usepackage{eucal}
\usepackage{amsmath}
\usepackage{amsthm}

\allowdisplaybreaks

\def\beq{\begin{equation}}
\def\eeq{\end{equation}}
\def\bea{\begin{eqnarray}}
\def\eea{\end{eqnarray}}
\let\nn=\nonumber
\def\beann{\begin{eqnarray*}}
\def\eeann{\end{eqnarray*}}

\let\a=\alpha \let\be=\beta \let\g=\gamma \let\de=\delta
  \let\h=\eta 

 \let\k=\kappa  
 \let\x=\xi   \let\s=\sigma
 
 \let\Ph=\phi  
  
  \let\D=\Delta

\let\qd=\quad \let\qqd=\qquad \def\qqqd{\qquad\qquad}

\def\tst#1{{\textstyle #1}}

\def\sst#1{{\scriptstyle #1}}

\theoremstyle{plain}

\newtheorem{lemma}{Lemma}

\newtheorem*{corollary*}{Corollary}

\theoremstyle{definition}
\newtheorem*{remark}{Remark}

\def\2{\frac{1}{2}} \def\4{\frac{1}{4}}

\def\6{\partial}

\def\<{\langle} \def\>{\rangle}

\let\auf=\uparrow \let\ab=\downarrow

  \def\CL{{\mathcal L}}
 \def\CT{{\mathcal T}} \def\CO{{\mathcal O}}
\def\CR{{\mathcal R}}

\def\i{{\rm i}}

\def\ctg{{\rm ctg}}

\def\sh{{\rm sh}}

\def\sn{{\rm sn}}  

 \def\End{{\rm End}} \def\id{{\rm id}}

\def\tr{{\rm tr}} 
\def\str{{\rm str}}

\begin{document}

\thispagestyle{empty}

\begin{center}
{\Large {\bf Solution of the quantum inverse problem\\}}
\vspace{7mm}
{\large F.~G\"{o}hmann\footnote[2]
{e-mail: goehmann@insti.physics.sunysb.edu} and
V.~E.~Korepin\footnote[1]{e-mail: korepin@insti.physics.sunysb.edu}}\\
\vspace{5mm}
C. N. Yang Institute for Theoretical Physics,\\ State University of New
York at Stony Brook,\\ Stony Brook, NY 11794-3840, USA\\
\vspace{20mm}

{\large {\bf Abstract}}
\end{center}
\begin{list}{}{\addtolength{\rightmargin}{10mm}
               \addtolength{\topsep}{-5mm}}
\item
We derive a formula that expresses the local spin and field
operators of fundamental graded models in terms of the elements of the
monodromy matrix. This formula is a quantum analogue of the classical
inverse scattering transform. It applies to fundamental spin chains,
such as the XYZ chain, and to a number of important exactly solvable
models of strongly correlated electrons, such as the supersymmetric
$t$-$J$ model or the the EKS model.
\\[2ex]
\end{list}

\clearpage

\section*{Introduction}
The quantum inverse scattering method was initiated some twenty years
ago by E. K. Sklyanin and L. D. Faddeev \cite{SkFa78a,SkFa78b} and then
mostly developed by the group of the Steklov mathematical institute at
Leningrad (cf, for instance, \cite{TaFa79,Sklyanin80,KuSk80,FaTa81}).
A pedagogical account of its most important aspects can be found in
\cite{KBIBo}.

The quantum inverse scattering method got its name, since it arose as
an attempt to develop a quantum version of the (classical) inverse
scattering method \cite{AbSe81,FaTa87}, which was successful in solving
non-linear classical evolution equations, such as the Korteweg-de-Vries
equation \cite{GGKM67}, the non-linear Schr\"odinger equation
\cite{ZhSh71} or the sine-Gordon equation \cite{AKNS73}.

The classical inverse scattering method provides a mapping from a set
of field variables satisfying non-linear evolution equations to a set
of scattering data of an associated auxiliary problem. While the
fields obey non-linear evolution equations, the scattering data evolve
linearly in time. The solution of the initial value problem for the
original non-linear evolution equations of the fields is achieved by
first mapping the initial data to the scattering data at time $t = 0$,
then using the linear time evolution of the scattering data, and
finally applying the {\em inverse transformation} \cite{GeLe51,%
Marchenko55} from scattering data to fields at a time $t > 0$.

In this paper we solve the `inverse scattering problem' for quantum
lattice models. The solution is remarkably simple.

Nowadays the term `quantum inverse scattering method' usually
refers to a method formulated for systems of finite length. The relation
to the classical case is the following. The elements of the monodromy
matrix, which  appears in the formulation of the classical
problem for systems of finite length, have simple Poisson brackets
\cite{Sklyanin80}. In the quantum case the Poisson brackets are
replaced by commutators of quantum operators. These commutators remain
simple after quantization. The quantum operators can be grouped into a
matrix, which, by analogy to the classical case, is called (quantum)
monodromy matrix. The elements of the quantum monodromy matrix obey a
set of quadratic relations. They generate the so-called Yang-Baxter
algebra. The structure of this algebra is determined by numerical
functions of a complex spectral parameter, which again can be arranged
in a matrix. This matrix is called the $R$-matrix. It satisfies the
famous Yang-Baxter equation (see (\ref{ybe}) below). The $R$-matrix
and its associated Yang-Baxter algebra are the key concepts of the
quantum inverse scattering method. These concepts are {\em algebraic}.

The Yang-Baxter algebra has two primary applications. First of all,
it contains, in general, a rich commutative subalgebra generated by
the trace of the monodromy matrix. The elements of this subalgebra
have a natural interpretation as a set of commuting quantities
belonging to a physical system. One of these quantities is interpreted
as the Hamiltonian. The existence of a large set of commuting operators
can not be directly utilized to diagonalize the Hamiltonian. In many
cases, however, the Yang-Baxter algebra can be employed for this task.
It can be used to simultaneously diagonalize all of the commuting
quantities by a procedure called the algebraic Bethe ansatz
\cite{TaFa79}. This is the most important application of the Yang-Baxter
algebra.

Inspite of the conceptual differences between classical and quantum
inverse scattering method, both methods have an important point in
common. They  essentially rely on a mapping from local field variables
to a set of non-local variables, which are the elements of the
monodromy matrix. In the quantum case the inverse transformation,
expressing the local fields in terms of the elements of the monodromy
matrix, was not known until recently. It first appeared in the examples
of the inhomogeneous XXX and XXZ spin-$\frac{1}{2}$ Heisenberg chains
in the article \cite{KMT99a}. 

The article \cite{KMT99a} is part of a series of articles
\cite{MaSa96,KMT99a,IKMT99,KMT99pp} by Izergin, Kitanine, Maillet,
Sanches de Santos and Terras. In this series an interesting new
device, the `factorizing $F$-matrix' \cite{MaSa96} was introduced into
the algebraic Bethe ansatz and its features were explored.
This led to simplified derivations of a number of important results
for the XXX and XXZ spin-$\frac{1}{2}$ chains. Among the rederived
results are the norm formulae \cite{Korepin82,Gaudin83} and the Slavnov
formula \cite{Slavnov89} for the scalar product of a Bethe ansatz
eigenstate with a non-eigenstate. The articles \cite{KMT99a,IKMT99,%
KMT99pp} also provide simplified derivations of various results for
form factors \cite{KKS97} and correlation functions \cite{JMMN92,%
JiMi95,KIEU94} and their generalization to the case of non-zero
magnetic field.

{F}rom our point of view, the most interesting new result in
\cite{MaSa96,KMT99a,IKMT99,KMT99pp} is the solution of the quantum
inverse problem for the periodic,and inhomogeneous XXX and XXZ
spin-$\frac{1}{2}$ Heisenberg chains (cf \cite{KMT99a}). This result
appeared to be the most important new tool used in the rederivation of
the determinant formulae for form factors \cite{KMT99a} and in the
derivation of multiple integral representations of correlation
functions at finite magnetic field \cite{KMT99pp}.

In this article we shall focus on the quantum inverse problem. We shall
obtain an explicit solution, which is valid (i) in the homogeneous case%
\footnote{The proof of \cite{KMT99a} does not work in the homogeneous
case, since the $F$-matrices are not invertible for homogeneous
lattices.}, (ii) for models with $R$-matrices of arbitrary higher
dimension, and (iii), most generally, for fundamental graded models
\cite{GoMu98}. Upon specification to the cases of the inhomogeneous
XXX and XXZ spin-$\frac{1}{2}$ Heisenberg chains our result reduces to
the formula obtained in \cite{KMT99a}.

Our result in its most general form is given by the formula
(\ref{invers}) below. This formula is valid for homogeneous as well as
for inhomogeneous models. Important special cases considered in this
article are the solution of the quantum inverse problem for the
translationally invariant XYZ spin chain (cf equations
(\ref{inversxyz1})-(\ref{inversxyz3}) in section 1) and for the
inhomogeneous $t$-$J$ model (equations (\ref{tj1})-(\ref{tj8}) in
section 6).

Formula (\ref{invers}) expresses the local operators as {\em products}
of the entries of the monodromy matrix evaluated at the
inhomogeneities. The structure of the solution of the quantum inverse
problem for periodic lattice models is thus much different from the
structure of the solution of the classical inverse scattering problem.
In the quantum case we have an explicit multiplicative formula. In the
classical case the solution is implicit and additive. It reduces to
the Gelfand-Levitan-Marchenko integral equations \cite{GeLe51,%
Marchenko55}.

The article is organized as follows. In section 1 we present the
solution of the quantum inverse problem for the homogeneous XYZ
spin-$\frac{1}{2}$ chain. Later this solution will appear as a special
case of our general solution (\ref{invers}). We treat the case of the
XYZ chain separately, since the proof greatly simplifies in the
homogeneous case. In section 2 we remind to the reader the definitions
of graded vector spaces and graded associative algebras. We introduce
the notion of graded local projection operators which were recently
defined in \cite{GoMu98}. Section 3 reviews the construction of the
Yang-Baxter algebra for fundamental graded models \cite{GoMu98} and
some important results about the graded version of the quantum inverse
scattering method \cite{KuSk80,Kulish85,GoMu98}. In section 4 we
introduce canonical Fermi operators into the formalism. In section 5 we
illustrate the abstract formalism developed in the preceding section
through two examples which are important in physical applications.
We consider the small polaron model \cite{HiFr83} and the
supersymmetric $t$-$J$ model \cite{BrRi70,ZhRi88,Sutherland75,%
Schlottmann87,BaBl90,Sarkar91,EsKo92,FoKa93}. In section 6 we present
our main result (\ref{invers}) in its most general form, valid for
inhomogeneous, fundamental graded models associated with $R$-matrices
of arbitrary dimension. We specify our formula for the examples
considered in section 5 and work out its homogeneous limit. In section
7 we give a general definition of the fermionic $R$-operator associated
with a fundamental graded representation of the Yang-Baxter algebra.
A fermionic $R$-operator was recently introduced in \cite{USW98a,USW98b}
for a number of models important in physical applications. The
fermionic $R$-operator is one of the tools we shall need for the proof
of the solution (\ref{invers}) of the quantum inverse problem. Section
8 is devoted to this proof. We construct the shift operator for
inhomogeneous, fundamental graded models and work out its properties.
Our proof of the quantum invers problem solely relies on the
properties of the shift operator. In particular, we do not use
factorizing $F$-matrices (which are so far known only for the XXX and
XXZ spin-$\frac{1}{2}$ chains) as in the article \cite{KMT99a}. This
makes our approach more general and powerful. Our article is concluded
with a brief summary and a discussion of our new formulae.

\section{Solution of the quantum inverse problem for the XYZ chain}
In this section we shall start solving the quantum inverse problem
for fundamental models by considering an important example. The result
of this section will later appear as a special case of our main result,
equation (\ref{invers}). We think, however, that the structure of
our main result and of its proof is best understood by considering an
example first. We shall assume that the reader is familiar with the
basic ideas of the quantum inverse scattering method. Readers not
familiar with those ideas are referred to section 3, where a brief
review is provided.

The XYZ spin-$\frac{1}{2}$ chain is characterized by its $R$-matrix
\cite{Baxter72,Babook,TaFa79}
\begin{equation} \label{rxyz}
     R(u) = \begin{pmatrix}
		 a(u) & 0 & 0 & d(u) \\
		 0 & b(u) & c(u) & 0 \\
		 0 & c(u) & b(u) & 0 \\
		 d(u) & 0 & 0 & a(u)
              \end{pmatrix} \qd.
\end{equation}
In a normalization, which assures the unitarity of the $R$-matrix (cf
(\ref{unitary})), the Boltzmann weights $a(u), \dots, d(u)$ are given by
\begin{equation}
   \begin{aligned}
     a(u) & = \frac{\sn(u + 2\h)}{\sn(u) + \sn(2\h)} \qd, \\
     c(u) & = \frac{\sn(2\h)}{\sn(u) + \sn(2\h)} \qd,
   \end{aligned}
   \begin{aligned}
     \qd b(u) & = \frac{\sn(u)}{\sn(u) + \sn(2\h)} \qd, \\
         d(u) & = \frac{k \, \sn(u) \sn(2\h) \sn(u + 2\h)}
		       {\sn(u) + \sn(2\h)} \qd.
   \end{aligned}
\end{equation}
The $R$-matrix $R(u)$ is considered as acting on the tensor product
$\mathbb{C}^2 \otimes \mathbb{C}^2$. $R(u,v) := R(u - v)$ is a solution
of the Yang-Baxter equation (\ref{ybe}) (see below). Hence, an
exactly solvable spin chain can be associated with $R(u)$. Let us
briefly recall the steps necessary for its construction.

Define $e_\a^\be \in \End(\mathbb{C}^2)$, $\a, \be = 1, 2$ by
\begin{equation}
     e_1^1 = \tst{\begin{pmatrix} 1 & 0 \\ 0 & 0 \end{pmatrix}} \qd, \qd
     e_1^2 = \begin{pmatrix} 0 & 1 \\ 0 & 0 \end{pmatrix} \qd, \qd
     e_2^1 = \begin{pmatrix} 0 & 0 \\ 1 & 0 \end{pmatrix} \qd, \qd
     e_2^2 = \begin{pmatrix} 0 & 0 \\ 0 & 1 \end{pmatrix} \qd.
\end{equation}
The set $\{ e_\a^\be \in \End(\mathbb{C}^2) | \a, \be = 1, 2\}$ is a
basis of $\End(\mathbb{C}^2)$. The definition
\begin{equation} \label{defej22}
     {e_j}_\a^\be = I_2^{\otimes (j-1)} \otimes e_\a^\be \otimes
		    I_2^{\otimes (L-j)} \qd,
\end{equation}
for $j = 1, \dots, L$, and $I_2$ being the $2 \times 2$ unit matrix,
provides a basis of $\End(\mathbb{C}^2)^{\otimes L}$, which is the
space of states of an $L$-site spin-$\frac{1}{2}$ quantum spin chain.
The matrices ${e_j}_\a^\be$ satisfy
\begin{equation}
   [{e_j}_\a^\be, {e_k}_\g^\de] = 0 \qd \text{for $j \ne k$} \qd, \qd
   {e_j}_\a^\be {e_j}_\g^\de = \de^\be_\g {e_j}_\a^\de
\end{equation}
and have the meaning of local projection operators. Using the
$R$-matrix (\ref{rxyz}) and the local projection operators we can
define the $L$-matrix at site $j$ as
\begin{equation} \label{lxyz}
     L_j (u) = \sum_{\a, \be, \g, \de = 1}^2 \! R^{\a \be}_{\g \de} (u)
	       \, e_\a^\g \otimes {e_j}_\be^\de
             = \begin{pmatrix}
		      a(u) {e_j}_1^1 + b(u) {e_j}_2^2 &
		      c(u) {e_j}_2^1 + d(u) {e_j}_1^2 \\
		      d(u) {e_j}_2^1 + c(u) {e_j}_1^2 &
		      b(u) {e_j}_1^1 + a(u) {e_j}_2^2
                   \end{pmatrix} \, .
\end{equation}
The $L$-matrix $L_j (u)$ is a $2 \times 2$ matrix in an auxiliary space.
Its entries are operators acting on the space of states of an $L$-site
spin-$\frac{1}{2}$ chain. The monodromy matrix of the corresponding
homogeneous spin chain is the $L$-fold ordered product
\begin{equation} \label{monoxyz}
     T(u) = L_L (u) \dots L_1 (u) \qd.
\end{equation}
By construction the monodromy matrix gives a representation of the
Yang-Baxter algebra with $R$-matrix $R(u)$ (see section 3 below).
We have
\begin{equation} \label{ybaxyz}
     P R(u - v) \bigl( T(u) \otimes T(v) \bigr) =
	\bigl( T(v) \otimes T(u) \bigr) P R(u - v) \qd,
\end{equation}
where $P = \sum_{\a, \be = 1}^2 e_\a^\be \otimes e_\be^\a$ is the
permutation matrix on $\mathbb{C}^2 \otimes \mathbb{C}^2$. The
Yang-Baxter algebra (\ref{ybaxyz}) is the basis for the solution
of the XYZ chain by algebraic Bethe ansatz \cite{TaFa79}.

The monodromy matrix (\ref{monoxyz}) is a $2 \times 2$ matrix in
auxiliary space. Its entries are non-local operators acting on the
space of states of the XYZ chain. We may write the monodromy matrix as
\begin{equation}
     T(u) = \begin{pmatrix} A(u) & B(u) \\ C(u) & D(u) \end{pmatrix}
	    \qd.
\end{equation}
The quantum inverse problem is to express the local operators
${e_n}_\a^\be$ in terms of the elements $A(u), \dots, D(u)$ of the
monodromy matrix.

For the case at hand this problem is rather easily solved. Note that
$a(0) = c(0) = 1$ and $b(0) = d(0) = 0$. It follows from (\ref{lxyz})
that
\begin{equation} \label{lxyzzero}
     L_j (0) = \begin{pmatrix} {e_j}_1^1 & {e_j}_2^1 \\
		  {e_j}_1^2 & {e_j}_2^2 \end{pmatrix}
             = \sum_{\a, \be = 1}^2 e_\a^\be \otimes {e_j}_\be^\a
	     = P_{0j} \qd.
\end{equation}
Here $P_{0j}$ is the permutation operator that interchanges the
auxiliary space with the $j$-th quantum space. Similarly,
$P_{jk} = \sum_{\a, \be = 1}^2 {e_j}_\a^\be {e_k}_\be^\a$ interchanges
the $j$-th and $k$-th quantum spaces. Using equation (\ref{lxyzzero})
in equation (\ref{monoxyz}) we obtain
\begin{equation} \label{ridi}
     T(0) = P_{0L} P_{0 L-1} \dots P_{01}
	  = P_{01} P_{1L} P_{1 L-1} \dots P_{12}
	  = P_{01} \hat U \qd.
\end{equation}
Here $\hat U = P_{1L} P_{1 L-1} \dots P_{12} = P_{12} P_{23} \dots
P_{L-1 L}$ is the cyclic shift operator in quantum space. In the
second equation in (\ref{ridi}) we have used the commutation relation
$P_{jk} P_{jl} = P_{kl} P_{jk}$ for permutation operators. Let us write
equation (\ref{ridi}) in matrix form,
\begin{equation}
    \begin{pmatrix} A(0) & B(0) \\ C(0) & D(0) \end{pmatrix} =
    \begin{pmatrix} {e_1}_1^1 \, \hat U & {e_1}_2^1 \, \hat U \\
       {e_1}_1^2 \, \hat U & {e_1}_2^2 \, \hat U \end{pmatrix} \qd.
\end{equation}
Comparing the matrix elements we find the relations
\begin{align} \label{uad}
     \hat U & = A(0) + D(0) \qd, \\
     \s_1^- & = {e_1}_2^1 = B(0) \, \hat U^{-1} \qd, \\
     \s_1^+ & = {e_1}_1^2 = C(0) \, \hat U^{-1} \qd, \\
     \s_1^z & = {e_1}_1^1 - {e_1}_2^2 = \bigl( A(0) - D(0) \bigr) \,
		\hat U^{-1} \qd.
\end{align}
These relations constitute a solution of the quantum inverse problem
for the local operators acting on the first lattice site. We may now
simply use the shift operator to shift the site indices. Since
$\hat U^{n-1} {e_1}_\a^\be \hat U^{1-n} = {e_n}_\a^\be$ and $\hat U^L =
\id$, we obtain
\begin{align} \label{inversxyz1}
     \s_n^- & = \hat U^{n-1} B(0) \, \hat U^{L-n} \qd, \\
     \s_n^+ & = \hat U^{n-1} C(0) \, \hat U^{L-n} \qd, \\
     \s_n^z & = \hat U^{n-1} \bigl( A(0) - D(0) \bigr) \,
		\hat U^{L-n} \qd. \label{inversxyz3}
\end{align}
Taking into account equation (\ref{uad}) we see that the right hand
side of the equations (\ref{inversxyz1})-(\ref{inversxyz3}) are
entirely expressed in terms of the entries of the monodromy matrix.
An alternative way of writing (\ref{inversxyz1})-(\ref{inversxyz3}) is
\begin{equation} \label{inversxyze}
     {e_n}_\a^\be = \hat U^{n-1} \,  T_\a^\be (0) \, \hat U^{L-n}
		  = \bigl( A(0) + D(0) \bigr)^{n-1} \,  T_\a^\be (0) \,
		    \bigl( A(0) + D(0) \bigr)^{L-n} \qd.
\end{equation}
Equation (\ref{inversxyze}) allows us to calculate expectation values
of {\em local operators} by means of the Yang-Baxter algebra.

The remainder of this article will be devoted to the generalization of
equation (\ref{inversxyze}) to (i) arbitrary dimension of the
$R$-matrix, (ii) the inhomogeneous case, and (iii) to fundamental
graded models. It is important noting that our above solution of the
quantum inverse problem does not depend on the specific features of
the XYZ chain. Our calculation solely relied on the fact that the
$L$-matrix evaluated at $u = 0$ turns into a permutation operator
(cf equation (\ref{lxyzzero})).

\section{Graded vector spaces}
In this section we shall recall the basic concepts of graded vector
spaces and graded associative algebras. In the context of the quantum
inverse scattering method these concepts were first used by Kulish and
Sklyanin \cite{KuSk80,Kulish85}. We shall further recall the notions of
`graded local projection operators' and graded permutation operators.
Graded local projection operators were introduced in the article
\cite{GoMu98}. They enable the definition of fundamental graded
representations of the Yang-Baxter algebra, which will be given in the
following section.

Graded vector spaces are vector spaces equipped with a notion of odd
and even, that allows us to treat fermions within the formalism of the
quantum inverse scattering method. Let us start with a finite
dimensional local space of states $V$, on which we impose an
additional structure, the parity, from the outset. Let $V = V_0
\oplus V_1$, $\dim V_0 = m$, $\dim V_1 = n$. We shall call $v_0 \in
V_0$ even and $v_1 \in V_1$ odd. The subspaces $V_0$ and $V_1$ are
called the homogeneous components of $V$. The parity $p$ is a function
$V_i \rightarrow \mathbb{Z}_2$ defined on the homogeneous components
of~$V$,
\begin{equation}
     p(v_i) = i \qd, \qd i = 0, 1 \qd, \qd v_i \in V_i \qd.
\end{equation}
The vector space $V$ endowed with this structure is called a graded
vector space or super space. Let us fix a basis $\{e_1, \dots,
e_{m + n}\}$ of definite parity and let us define $p(\a) := p(e_\a)$.

The use of graded vector spaces within the quantum inverse scattering
method requires the construction of an algebra of commuting and
anticommuting {\it operators}. For this purpose we have to extend
the concept of parity to operators in End($V$) and to tensor products
of these operators. Let $e_\a^\be \in \End (V)$, $e_\a^\be e_\g =
\de_\g^\be e_\a$. The set $\{e_\a^\be \in \End (V) | \a, \be = 1,
\dots m + n \}$ is a basis of End($V$). Hence, the definition
\begin{equation} \label{gradend}
     p(e_\a^\be) = p(\a) + p(\be)
\end{equation}
induces a grading on End($V$) regarded as a vector space.

It is easy to see that an element $A = A^\a_\be e_\a^\be \in \End (V)$
is homogeneous with parity $p(A)$, if and only if
\begin{equation} \label{parend}
     (-1)^{p(\a) + p(\be)} A^\a_\be = (-1)^{p(A)} A^\a_\be \qd.
\end{equation}
The latter equation implies for two homogeneous elements $A, B \in
\End (V)$ that their product $AB$ is homogeneous with parity
\begin{equation} \label{homab}
     p(AB) = p(A) + p(B) \qd.
\end{equation}
In other words, multiplication of matrices in $\End (V)$ preserves
homogeneity, and, therefore, $\End (V)$ endowed with the grading
(\ref{gradend}) is a graded associative algebra \cite{KuSk80}.

Let us consider the $L$-fold tensorial power $(\End (V))^{\otimes L}$
of $\End (V)$. The definition (\ref{gradend}) has a natural extension
to $(\End (V))^{\otimes L}$, namely,
\begin{equation} \label{gradendl}
     p (e_{\a_1}^{\be_1} \otimes \dots \otimes e_{\a_L}^{\be_L})
          = p(\a_1) + p(\be_1) + \dots + p(\a_L) + p(\be_L) \qd.
\end{equation}
{F}rom this formula it can be seen in a similar way as before, that
homogeneous elements $A = A^{\a_1 \dots \a_L}_{\be_1 \dots \be_L}
e_{\a_1}^{\be_1} \otimes \dots \otimes e_{\a_L}^{\be_L}$ of
$(\End (V))^{\otimes L}$ with parity $p(A)$ are characterized by the
equation
\begin{equation}
     (-1)^{\sum_{j=1}^L (p(\a_j) + p(\be_j))}
        A^{\a_1 \dots \a_L}_{\be_1 \dots \be_L} =
     (-1)^{p(A)} A^{\a_1 \dots \a_L}_{\be_1 \dots \be_L} \qd,
\end{equation}
which generalizes (\ref{parend}). Again the product $AB$ is homogeneous
with parity $p(AB) = p(A) + p(B)$, if $A$ and $B$ are homogeneous. Thus
the definition (\ref{gradendl}) induces the structure of a graded
associative algebra on $(\End (V))^{\otimes L}$.

Let us define the super-bracket
\begin{equation} \label{superbracket}
     [X,Y]_\pm = XY - (-1)^{p(X)p(Y)} YX
\end{equation}
for $X$, $Y$ taken from the homogeneous components of End($V$), and let
us extend it linearly to $\End (V)$ in both of its arguments. Then,
End($V$) endowed with the super-bracket becomes the Lie-super algebra
gl($m|n$). Note that the above definition of a super-bracket makes
sense in any graded algebra and is particularly valid in $(\End
(V))^{\otimes L}$.

The following definition of `graded local projection operators'
\cite{GoMu98} will be crucial for our definition of fundamental graded
representations of the Yang-Baxter algebra in the next section.
Define the matrices
\begin{equation} \label{defej}
     {e_j}_\a^\be = (-1)^{(p(\a) + p(\be)) \sum_{k = j + 1}^L p(\g_k)}
		    \, I_{m + n}^{\otimes (j - 1)} \otimes e_\a^\be
		    \otimes e_{\g_{j+1}}^{\g_{j+1}} \otimes \dots
		    \otimes e_{\g_L}^{\g_L} \qd,
\end{equation}
where $I_{m + n}$ is the $(m + n) \times (m + n)$ unit matrix, and
summation over double {\em tensor indices} (i.e.\ over $\g_{j+1}, \dots,
\g_L$) is understood. We shall keep this sum convention throughout
the remainder of this article. The index $j$ on the left hand side of
(\ref{defej}) will later refer to the $j$-th site of a physical lattice
model and is called the site index. A simple consequence of the
definition (\ref{defej}) for $j \ne k$ are the commutation relations
\begin{equation} \label{coantico}
     {e_j}_\a^\be {e_k}_\g^\de = (-1)^{(p(\a) + p(\be))(p(\g) + p(\de))}
          {e_k}_\g^\de {e_j}_\a^\be \qd.
\end{equation}
It further follows from equation (\ref{defej}) that ${e_j}_\a^\be$ is
homogeneous with parity
\begin{equation}
     p({e_j}_\a^\be) = p(\a) + p(\be) \qd.
\end{equation}
Hence, in agreement with intuition, equation (\ref{coantico}) says that
odd matrices with different site indices mutually anticommute, whereas
even matrices commute with each other as well as with the odd matrices.
For products of matrices ${e_j}_\a^\be$ which are acting on the same
site (\ref{defej}) implies the projection property
\begin{equation} \label{samesite}
     {e_j}_\a^\be {e_j}_\g^\de = \de_\g^\be {e_j}_\a^\de \qd.
\end{equation}
Equations (\ref{coantico}) and (\ref{samesite}) justify our
terminology. The ${e_j}_\a^\be$ are graded analogues of local
projection operators. We call them graded local projection operators
or projection operators, for short. Using the super-bracket
(\ref{superbracket}), equations (\ref{coantico}) and (\ref{samesite})
can be combined into
\begin{equation} \label{imbglmn}
     [{e_j}_\a^\be,{e_k}_\g^\de]_\pm =
	  \de_{jk} \left( \de_\g^\be {e_j}_\a^\de -
	  (-1)^{(p(\a) + p(\be))(p(\g) + p(\de))}
          \de_\a^\de {e_j}_\g^\be \right) \qd.
\end{equation} 
The right hand side of the latter equation with $j = k$ gives the
structure constants of the Lie super algebra gl($m|n$) with respect
to the basis $\{{e_j}_\a^\be\}$.

Since any $m + n$-dimensional vector space over the complex numbers
is isomorphic to $\mathbb{C}^{m+n}$, we may simply set $V =
\mathbb{C}^{m+n}$. We may further assume that our homogeneous basis
$\{ e_\a \in \mathbb{C}^{m+n}| \a = 1, \dots, m + n \}$ is canonical,
i.e.\ we may represent the vector $e_\a$ by a column vector having the
only non-zero entry +1 in row $\a$. Our basic matrices $e_\a^\be$ are
then $(m + n) \times (m + n)$-matrices with a single non-zero entry +1
in row $\a$ and column $\be$.

\begin{remark}
The meaning of (\ref{defej}) becomes more evident by considering a
simple example. Let $m = n = 1$ and $p(1) = 0$, $p(2) = 1$. Then, using
(\ref{imbglmn}), we obtain
\begin{align} \label{cogl111}
     [{e_j}_1^2, {e_k}_1^2]_\pm & =
	       \{ {e_j}_1^2, {e_k}_1^2 \} = 0 \qd, \\ \label{cogl112}
     [{e_j}_2^1, {e_k}_2^1]_\pm & =
	       \{ {e_j}_2^1, {e_k}_2^1 \} = 0 \qd, \\
     [{e_j}_1^2, {e_k}_2^1]_\pm & = 
	       \{ {e_j}_2^1, {e_k}_1^2 \} = \de_{jk}
	          ({e_j}_1^1  + {e_j}_2^2) = \de_{jk}
\end{align}
for $j, k = 1, \dots, L$. The curly brackets in (\ref{cogl111}),
(\ref{cogl112}) denote the anticommutator. The matrices ${e_j}_1^2$
and ${e_k}_2^1$ satisfy the canonical anticommutation relations for
spinless Fermi operators. We can therefore identify ${e_j}_1^2
\rightarrow c_j$ and ${e_k}_2^1 \rightarrow c_k^+$. Introducing Pauli
matrices $\s^+ = e_1^2$, $\s^- = e_2^1$ and $\s^z = e_1^1 - e_2^2$ we
obtain, by carrying out the summation, the following explicit matrix
representation from our basic definition (\ref{defej}),
\begin{align}
     c_j & = I_2^{\otimes (j - 1)} \otimes \s^+ \otimes
	     (\s^z)^{\otimes (L - j)} \qd, \\
     c_k^+ & = I_2^{\otimes (k - 1)} \otimes \s^- \otimes
	       (\s^z)^{\otimes (L - k)} \qd.
\end{align}
This is the well-known Jordan-Wigner transformation \cite{JoWi28}
expressing Fermi operators for spinless fermions in terms of Pauli
matrices. We may thus interpret equation (\ref{defej}) as a
generalization of the Jordan-Wigner transformation. In general,
equation (\ref{defej}) provides matrix representations not of Fermi
operators but, more generally, of fermionic projection operators.
Representations of Fermi operators can be obtained be taking
appropriate linear combinations of matrices ${e_j}_\a^\be$. This point
will be elaborated in section 4 below. Note that an alternative
generalization of the Jordan-Wigner transformation, which relies on
representations of the Clifford algebra, was recently introduced
in~\cite{DaMa98}.
\end{remark}

The permutation operator plays an important role in the construction of 
local integrable lattice models. It enters the expression for the shift
operator on homogeneous lattices and the expression for the Hamiltonian
of rational gl($m|n$) invariant models (see section 5 below). In the
graded case the definition of the permutation operator requires the
following modifications of signs,
\begin{equation} \label{defp}
     P_{jk} = (-1)^{p(\be)} {e_j}_\a^\be {e_k}_\be^\a \qd.
\end{equation}
As indicated by its name, this operator induces the action of the
symmetric group $\mathfrak{S}^L$ on the site indices of the matrices
${e_j}_\a^\be$. The properties of $P_{jk}$ (for $j \ne k$) are the
same as in the non-graded case. They are easily derived from
(\ref{coantico}) and (\ref{samesite}) and can be found, for instance,
in \cite{GoMu98}. Let $L = 2$. Then
\begin{equation}
     P_{12} = (-1)^{p(\be)} {e_1}_\a^\be {e_2}_\be^\a
            = (-1)^{p(\a)p(\be)} e_\a^\be \otimes e_\be^\a
	    = (-1)^{p(\a)p(\be)} \de_\de^\a \de_\g^\be
	      e_\a^\g \otimes e_\be^\de \qd.
\end{equation}
{F}rom the right hand side of this equation we can read off the matrix
elements of $P_{12}$ with respect to the canonical basis of
End($V \otimes V$).
\section{Fundamental graded models}
In this section we shall recall the notion of {\em fundamental graded
representations} of the Yang-Baxter algebra, which was recently
introduced in \cite{GoMu98}. For a given grading we shall associate
a fundamental model with every solution of the Yang-Baxter equation that
satisfies a certain compatibility condition (see (\ref{comp}) below).

\unitlength .85mm
\begin{figure}

\begin{picture}(140,50)


\put(12,15){\line(1,0){13}}
\put(12,30){\line(1,0){13}}
\put(25,15){\line(1,1){15}}
\put(25,30){\line(1,-1){15}}
\put(40,15){\vector(1,0){25}}
\put(40,30){\vector(1,0){25}}
\put(51,37){\vector(0,-1){32}}

\put(14,10){$u$}
\put(14,25){$v$}
\put(46,33){$w$}

\put(71.5,21.5){=}

\put(122,15){\vector(1,0){13}}
\put(122,30){\vector(1,0){13}}
\put(107,15){\line(1,1){15}}
\put(107,30){\line(1,-1){15}}
\put(107,15){\line(-1,0){25}}
\put(107,30){\line(-1,0){25}}
\put(96,37){\vector(0,-1){32}}

\put(84,10){$u$}
\put(84,25){$v$}
\put(91,33){$w$}

\end{picture}
\caption{The Yang-Baxter equation is most easily memorized in
graphical form}
\end{figure}
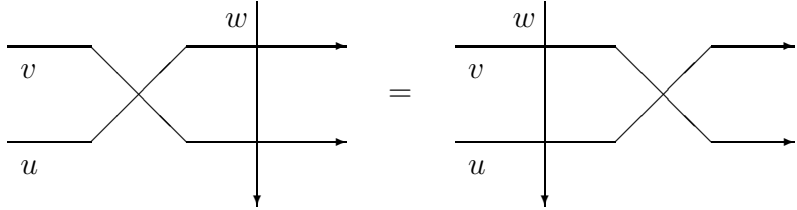

For our present purpose it is most suitable to interpret the Yang-Baxter
equation as a set of functional equations for the matrix elements
of an $(m + n)^2 \times (m + n)^2$-matrix $R(u,v)$. We may represent the
Yang-Baxter equation in graphical form as shown in Figure 1, where each
vertex corresponds to a factor in the equation
\begin{equation} \label{ybe}
     R_{\a' \be'}^{\a \be} (u,v) R_{\a'' \g'}^{\a' \g} (u,w)
     R_{\be'' \g''}^{\be' \g'} (v,w) =
     R_{\be' \g'}^{\be \g} (v,w) R_{\a' \g''}^{\a \g'} (u,w)
     R_{\a'' \be''}^{\a' \be'} (u,v) \qd .
\end{equation}
Note that there is a direction assigned to every line in figure 1,
which is indicated by the tips of the arrows. Therefore every vertex
has an orientation, and vertices and $R$-matrices can be identified
according to figure 2, where indices have been supplied to a vertex.
Summation is over all inner lines in figure 1.
\begin{figure}

\begin{picture}(140,40)

\put(25,20){$R^{\a \be}_{\g \de} (u,v) \qd =$}

\put(70,21.5){\vector(1,0){26}}
\put(82,33.5){\vector(0,-1){26}}

\put(71,17){$u$}
\put(78,30.5){$v$}

\put(64,21){$\sst{\a}$}
\put(81,38){$\sst{\be}$}
\put(81,1){$\sst{\de}$}
\put(100,21){$\sst{\g}$}

\end{picture}

\caption{Identification of the $R$-matrix with a vertex}
\end{figure}

The construction of a graded Yang-Baxter algebra and its fundamental
representation requires only minimal modifications compared to the
non-graded case \cite{KBIBo}. Let us assume we are given a solution of
(\ref{ybe}), which is compatible with the grading in the sense that
\cite{KuSk80}
\begin{equation} \label{comp}
     R_{\g \de}^{\a \be} (u,v) = (-1)^{p(\a) + p(\be) + p(\g) + p(\de)}
     R_{\g \de}^{\a \be} (u,v) \qd.
\end{equation}
Define a graded $L$-matrix  at site $j$ as
\begin{equation} \label{defgl}
     {\CL_j}^\a_\be (u,v) = (-1)^{p(\a) p(\g)}
         R^{\a \g}_{\be \de} (u,v) {e_j}_\g^\de \qd.
\end{equation}
Equation (\ref{comp}) implies that the matrix elements of ${\CL_j}
(u,v)$ are of definite parity,
\begin{equation}
     p( {\CL_j}^\a_\be (u,v)) = p(\a) + p(\be) \qd.
\end{equation}
Thus their commutation rules are given by
\begin{equation} \label{ljlk}
     {\CL_j}^\a_\be (u,v) {\CL_k}^\g_\de (w,z) =
          (-1)^{(p(\a) + p(\be))(p(\g) + p(\de))}
	  {\CL_k}^\g_\de (w,z) {\CL_j}^\a_\be (u,v) \qd.
\end{equation}
It further follows from the Yang-Baxter equation (\ref{ybe}) and
from equation (\ref{comp}) that
\begin{equation} \label{gyba}
     \check R(u,v) \bigl( \CL_j (u,w) \otimes_s \CL_j (v,w) \bigr) =
     \bigl( \CL_j (v,w) \otimes_s \CL_j (u,w) \bigr) \check R(u,v) \qd.
\end{equation}
As in the non-graded case the matrix $\check R (u,v)$ is defined by
\begin{equation} \label{rcheck}
     \check R^{\a \be}_{\g \de} (u,v) = R^{\be \a}_{\g \de} (u,v) \qd.
\end{equation}
The super tensor product \cite{KuSk80} in equation (\ref{gyba}) is to
be understood as a super tensor product of matrices with non-commuting
entries, $(A \otimes_s B)^{\a \g}_{\be \de} = (-1)^{(p(\a) + p(\be))
p(\g)} A^\a_\be B^\g_\de$. The super tensor product has the following
important feature. Given matrices $A$, $B$, $C$, $D$ with operator
valued entries, which mutually commute according to the same rule as
$\CL_j$ and $\CL_k$ in equation (\ref{ljlk}), we obtain for the product
of two super tensor products
\begin{equation} \label{abcd}
     (A \otimes_s B)(C \otimes_s D) = AC \otimes_s BD \qd.
\end{equation}
Equation (\ref{gyba}) may be interpreted as defining a graded
Yang-Baxter algebra with $R$-matrix $R$. We call $\CL_j$ its
fundamental graded representation.

Starting from (\ref{gyba}) we can construct integrable lattice models
as in the non-graded case \cite{KBIBo}. Let us briefly recall the
construction with emphasis on the modifications that appear due to the
grading. Define a monodromy matrix $\CT (u,v)$ as an $L$-fold ordered
product of fundamental $L$-matrices,
\begin{equation} \label{mono}
     \CT (u,v) = \CL_L (u,v) \dots \CL_1 (u,v) \qd.
\end{equation}
Due to equation (\ref{homab}) the matrix elements of $\CT(u,v)$ are
homogeneous with parity $p(\CT^\a_\be (u,v)) = p(\a) + p(\be)$. Repeated
application of (\ref{gyba}) and (\ref{abcd}) shows that this monodromy
matrix is a representation of the graded Yang-Baxter algebra,
\begin{equation} \label{gtyba}
     \check R(u,v) \bigl( \CT (u,w) \otimes_s \CT (v,w) \bigr) =
     \bigl( \CT (v,w) \otimes_s \CT (u,w) \bigr) \check R(u,v) \qd.
\end{equation}
In the non-graded case ($n = 0$) the super tensor product in
(\ref{gtyba}) agrees with the usual tensor product. Let us now define
the super trace as
\begin{equation}
     \str( A) = (-1)^{p(\a)} A^\a_\a \qd.
\end{equation}
It follows from (\ref{comp}) and (\ref{gtyba}) that
\begin{equation} \label{trans}
     \bigr[\str( \CT (u,w)),\str( \CT (v,w))\bigr] = 0 \qd,
\end{equation}
which is in complete analogy with the non-graded case.

Let us assume that $R(u,v)$ is a regular solution of the Yang-Baxter
equation, $R^{\a \be}_{\g \de} (v, v) = \de^\a_\de \de^\be_\g$. Then
(\ref{defgl}) implies that
\begin{equation}
     {\CL_j}^\a_\be (v, v) = (-1)^{p(\a) p(\be)} {e_j}^\a_\be \qd,
\end{equation}
and we can easily see \cite{GoMu98} that the super trace of the
monodromy matrix evaluated at $(v, v)$ generates a shift by one site,
\begin{equation} \label{shift}
     \str(\CT (v, v)) = P_{12} P_{23} \dots P_{L-1 L} =: \hat U \qd.
\end{equation}
It follows that $\tau (u) := - \i \ln(\str( \CT (u, v)))$ generates a
sequence of local operators \cite{Luescher76} which, as a consequence 
of (\ref{trans}), mutually commute,
\begin{equation} \label{tau}
     \tau(u) = \hat \Pi + (u - v) \hat H
			   + \CO \left( (u - v)^2 \right) \qd.
\end{equation}
$\hat \Pi$ in this expansion is the momentum operator. On a lattice,
where the minimal possible shift is by one site, and thus $\hat U$
rather than $\hat \Pi$ is the fundamental geometrical operator, some
care is required in the definition of $\hat \Pi$. As was shown in
\cite{GoMu97} a proper definition may be obtained by setting
$\Pi := - \i \ln ( \hat U) \mod 2 \pi$ and expressing the function
$f(x) = x \mod 2 \pi$ by its Fourier sum. Then $\hat \Pi$ becomes a
polynomial in $\hat U$.
\begin{equation}
     \hat \Pi = \Ph \sum_{m=1}^{L-1} \left( \2 +
		 \frac{\hat U^m}{e^{- \i \Ph m} - 1} \right) \qd,
\end{equation}
where $\Ph = 2 \pi /L$. The first order term $\hat H$ in the expansion
(\ref{tau}) may be interpreted as Hamiltonian. Using (\ref{shift}) it
is obtained as
\begin{equation} \label{ham}
     \hat H = \sum_{j=1}^L H_{j j+1} \qd,
\end{equation}
where $H_{L L+1} = H_{L 1}$ and
\begin{equation} \label{hdens}
     H_{j j+1} = - \i \, (-1)^{p(\g)(p(\a) + p(\g))} \,
                 \6_u \left. \check R^{\a \be}_{\g \de} (u,v)
		 \right|_{u = v} {e_j}_\a^\g {e_{j+1}}_\be^\de \qd.
\end{equation}

We would like to draw the readers' attention to the following points.
(i) The $R$-matrix $\check R$ in equation (\ref{gyba}) does {\it not}
undergo a modification due to the grading. (ii) The only necessary
compatibility condition which has to be satisfied in order to introduce
a fundamental graded representation of the Yang-Baxter algebra
associated with a solution of the Yang-Baxter equation is equation
(\ref{comp}), which was introduced in \cite{KuSk80}. Equation
(\ref{comp}) is a rather weak constraint. The most important solutions
of the Yang-Baxter equation, which appear in physical applications are
compatible with a non-trivial grading (see the examples in section 5).
Moreover, a given $R$-matrix may be compatible with different gradings,
leading to different fundamental graded representations of the
Yang-Baxter algebra \cite{GoMu98}.

Before turning to our next subject let us introduce the inhomogeneous
generalization
\begin{equation} \label{tin}
     \CT(u; \xi_1, \dots, \xi_L) = \CL_L (u, \xi_L)
				      \dots \CL_1 (u, \xi_1)
\end{equation}
of the monodromy matrix (\ref{mono}). This monodromy matrix satisfies
(\ref{gtyba}). For $\x_1 = \dots = \x_L = v$ it turns into $\CT(u,v)$
defined in (\ref{mono}). We shall formulate our main result below for
the inhomogeneous model generated by $\CT(u; \xi_1, \dots, \xi_L)$.
\section{Fermi operators}
In the article \cite{GoMu98} it was explained how the various graded
objects introduced in the previous section can be expressed in terms
of Fermi operators. The key observation is that, as far as the matrices
${e_j}_\a^\be$ are concerned, all calculations of the previous section
rely on the commutation relations (\ref{coantico}) and on the
projection property (\ref{samesite}). Fermionic projection operators
satisfy the same equations. We may thus say that the matrices
${e_j}_\a^\be$ are matrix representations of fermionic projection
operators. As we have seen in the previous section, the matrices
${e_j}_\a^\be$ are suitable for formulating a graded version of the
quantum inverse scattering method. For the physical interpretation
of the models constructed from a given solution of the Yang-Baxter
equation, however, it is convenient to introduce Fermi operators into
the formalism.

A general construction of fermionic projection operators for fermions
of arbitrary su(N) spin was presented in \cite{GoMu98}. Rather than
repeating those results let us illustrate them by example.

Let us consider spinless fermions on a ring of $L$ lattice sites,
\begin{equation} \label{antic}
     \{c_j,c_k\} = \{c_j^\dagger,c_k^\dagger\} = 0 \qd, \qd
     \{c_j,c_k^\dagger\} = \de_{jk} \qd, \qd j, k = 1, \dots , L \qd.
\end{equation}
It is easy to verify that the entries $(X_j)^\a_\be$ of the matrix
\begin{equation} \label{xj1}
     X_j = \begin{pmatrix}
	      1 - n_j & c_j \\ c_j^\dagger & n_j
	   \end{pmatrix}
\end{equation}
are fermionic projection operators. Define ${X_j}_\a^\be =
(X_j)^\a_\be$. Then
\begin{equation} \label{xpro}
     {X_j}_\a^\be {X_j}_\g^\de = \de_\g^\be {X_j}_\a^\de \qd.
\end{equation}
The operators ${X_j}_\a^\be$ carry parity, induced by the anti
commutation rule (\ref{antic}) for the Fermi operators. For $j \ne k$
${X_j}_\a^\be$ and ${X_k}_\g^\de$ anticommute, if both are build up of
an odd number of Fermi operators, and otherwise commute. This fact can
be expressed as follows. Let $p(1) = 0$, $p(2) = 1$ and
$p({X_j}_\a^\be) = p(\a) + p(\be)$. Then ${X_j}_\a^\be$ is odd
(contains an odd number of Fermi operators), if $p({X_j}_\a^\be) = 1$,
and even, if $p({X_j}_\a^\be) = 0$. The commutation rules for the
projectors ${X_j}_\a^\be$ are thus
\begin{equation} \label{xcom}
     {X_j}_\a^\be {X_k}_\g^\de = (-1)^{(p(\a) + p(\be))(p(\g) + p(\de))}
        {X_k}_\g^\de {X_j}_\a^\be \qd.
\end{equation}
Now (\ref{xpro}) and (\ref{xcom}) are of the same form as
(\ref{samesite}) and (\ref{coantico}), respectively. Since the
calculations in the previous section relied solely on (\ref{coantico})
and (\ref{samesite}), we may simply replace ${e_j}_\a^\be \rightarrow
{X_j}_\a^\be$ in equations (\ref{defgl}) and (\ref{hdens}).

Fermionic representations compatible with arbitrary grading can be
constructed by considering several species of fermions and graded
products of projection operators. We shall explain this for the case
of two species. This is the case most interesting for applications,
since we may interpret the two species as up- and down-spin electrons.
We have to attach a spin index to the Fermi operators,
$c_j \rightarrow c_{j \s}$, $\s = \auf, \ab$,
$\{c_{j \s},c_{k \tau}^\dagger\} = \de_{jk} \de_{\s \tau}$.
Accordingly, there are two species of projection operators,
${X_j}_\a^\be \rightarrow {X_j^\s}_\a^\be$.

Let us define projection operators for electrons by the tensor products
\begin{equation}
     {X_j}_{\a \g}^{\be \de} = (-1)^{(p(\a) + p(\be))p(\g)}
        {X_j^\ab}_\a^\be {X_j^\auf}_\g^\de =
	\left( X_j^\ab \otimes_s X_j^\auf \right)^{\a \g}_{\be \de} \qd.
\end{equation}
Then
\begin{equation}
     {X_j}_{\a \g}^{\be \de} {X_j}_{\a' \g'}^{\be' \de'} =
        \de_{\a'}^\be \de_{\g'}^\de {X_j}_{\a \g}^{\be' \de'} \qd.
\end{equation}
${X_j}_{\a \g}^{\be \de}$ inherits the parity from ${X_j^\ab}_\a^\be$
and ${X_j^\auf}_\g^\de$. The number of Fermi operators contained in
${X_j}_{\a \g}^{\be \de}$ is the sum of the number of Fermi operators
in ${X_j^\ab}_\a^\be$ and ${X_j^\auf}_\g^\de$. Hence
$p({X_j}_{\a \g}^{\be \de}) = p({X_j^\ab}_\a^\be) + p({X_j^\auf}_\g^\de)
= p(\a) + \dots + p(\de)$, and the analog of (\ref{xcom}) holds for
${X_j}_{\a \g}^{\be \de}$, too. Again we present all projection
operators in form of a matrix $(X_j)^{\a \g}_{\be \de} =
{X_j}_{\a \g}^{\be \de}$,
\bea
     X_j & = & X_j^\ab \otimes_s X_j^\auf \nn \\ \label{xj}
        & = &
        \begin{pmatrix}
	   (1 - n_{j \ab})(1 - n_{j \auf}) &
	   (1 - n_{j \ab}) c_{j \auf} &
	   c_{j \ab} (1 - n_{j \auf}) &
	   c_{j \ab} c_{j \auf} \\
	   (1 - n_{j \ab}) c_{j \auf}^\dagger &
	   (1 - n_{j \ab}) n_{j \auf} &
	   - c_{j \ab} c_{j \auf}^\dagger &
	   - c_{j \ab} n_{j \auf} \\
	   c_{j \ab}^\dagger (1 - n_{j \auf}) &
	   c_{j \ab}^\dagger c_{j \auf} &
	   n_{j \ab} (1 - n_{j \auf}) &
	   n_{j \ab} c_{j \auf} \\
	   - c_{j \ab}^\dagger c_{j \auf}^\dagger &
	   - c_{j \ab}^\dagger n_{j \auf} &
	   n_{j \ab} c_{j \auf}^\dagger &
	   n_{j \ab} n_{j \auf}
        \end{pmatrix} \,. \qd
\eea
Here we used the standard ordering of matrix elements of tensor
products, corresponding to a renumbering $(11) \rightarrow 1$,
$(12) \rightarrow 2$, $(21) \rightarrow 3$, $(22) \rightarrow 4$.
Within this convention ${X_j}_{\a \g}^{\be \de}$ is replaced by
${X_j}_\a^\be$, $\a, \be = 1, \dots, 4$, which then satisfies
(\ref{xpro}) and (\ref{xcom}) with grading $p(1) = p(4) = 0$,
$p(2) = p(3) = 1$.

Note that Fermi operators can be obtained as linear combinations of
projection operators. We have, for instance, $c_{j \auf} = {X_j}_1^2 +
{X_j}_3^4$.

So far we have considered the case of spinless fermions with
two-di\-men\-sional local space of states and grading $m = n =1$, and
the case of electrons with four-dimensional space of states and grading
$m = n = 2$. There are four different possibilities to realize
(\ref{coantico}) and (\ref{samesite}) in case of a three-dimensional
local space of states, $m + n = 3$. They can be obtained by deleting
the $\a$'s row and column of the matrix $X_j$ in equation (\ref{xj}),
$\a = 1, 2, 3, 4$. (\ref{xpro}) and (\ref{xcom}) remain valid, since
the operators ${X_j}_\a^\be$ are projectors.

An alternative way \cite{PuZh86,OWA87,DaMa98} of introducing Fermi
operators into the quantum inverse scattering method is by applying the
Jordan-Wigner transformation \cite{JoWi28} to the non-graded $L$-matrix
and then pulling out the non-local factors. This approach was of primary
importance, for instance, for a fermionic formulation of the
Yang-Baxter algebra of the Hubbard model \cite{OWA87} and led to the
discovery of a SO(4)-invariant form of the monodromy matrix of the
Hubbard model \cite{OWA87,GoMu97,SUW98}. In general, however, we prefer
the method presented above, since the approach of \cite{PuZh86,OWA87,%
DaMa98} so far did not lead to general formulae such as (\ref{defgl}) or
(\ref{hdens}) and may have unpleasant side effects, such as a twist of
boundary conditions or the appearance of numerous factors of `i' in
the equations.
\section{Examples}
Before turning to our main result let us present several examples in
order to provide an idea to the reader which applications we have in
mind.

We shall start with the XXZ spin chain. The $R$-matrix of the XXZ spin
chain can, for instance, be written as
\begin{equation} \label{rxxz}
     R(u,v) = \begin{pmatrix}
		 1 & 0 & 0 & 0 \\
		 0 & b(u,v) & c(u,v) & 0 \\
		 0 & c(u,v) & b(u,v) & 0 \\
		 0 & 0 & 0 & 1
              \end{pmatrix} \qd,
\end{equation}
where
\begin{equation} \label{defbc}
     b(u,v) = \frac{\sh(u - v)}{\sh(u - v + \i \k)} \qd, \qd
     c(u,v) = \frac{\sh(\i \k)}{\sh(u - v + \i \k)} \qd.
\end{equation}
The $R$-matrix (\ref{rxxz}) is compatible with the grading $p(1) = 0$,
$p(2) = 1$ (cf equation (\ref{comp})). The corresponding $L$-matrix
then follows from (\ref{defgl}),
\begin{equation}
     \CL_j (u,v) = \begin{pmatrix}
		      {e_j}_1^1 + b(u,v) {e_j}_2^2 & c(u,v) {e_j}_2^1 \\
		      c(u,v) {e_j}_1^2 & b(u,v) {e_j}_1^1 - {e_j}_2^2
                   \end{pmatrix} \qd.
\end{equation}
Using (\ref{ham}), (\ref{hdens}) we obtain the Hamiltonian
\begin{equation}
     \hat H = \frac{-1}{\sin(\k)} \sum_{j=1}^L \Bigl\{
		 {e_j}_2^1 {e_{j+1}}_1^2 + {e_{j+1}}_2^1 {e_j}_1^2 -
		 \cos (\k) \left(
		 {e_j}_1^1 {e_{j+1}}_2^2 + {e_{j+1}}_1^1 {e_j}_2^2
		 \right) \Bigr\} \qd.
\end{equation}
We may now replace the matrices ${e_j}_\a^\be$ by the fermionic
projectors ${X_j}_\a^\be$, equation (\ref{xj1}). Then
\begin{equation}
     \CL_j (u,v) = \begin{pmatrix}
		      (1 - n_j) + b(u,v) n_j & c(u,v) \, c_j^+ \\
		      c(u,v) \, c_j & b(u,v) (1 - n_j) - n_j
                   \end{pmatrix}
\end{equation}
and
\begin{equation} \label{smallp}
     \hat H = \frac{-1}{\sin(\k)} \sum_{j=1}^L \bigl\{
		 c_j^+ c_{j+1} + c_{j+1}^+ c_j + 2 \,\cos (\k)
		 \, n_j n_{j+1} \bigr\} + 2 \ctg(\k) \hat N \qd,
\end{equation}
where $\hat N = \sum_{j=1}^L n_j$ is the particle number operator.
The Hamiltonian (\ref{smallp}) defines the `small polaron model'
\cite{HiFr83}. Note that the algebraic Bethe ansatz for the small
polaron model gets slightly modified compared to the `non-graded' spin
chain case ($p(1) = p(2) = 0$), since due to the grading $p(1) = 0$,
$p(2) = 1$ there appear certain minus signs in the Yang-Baxter algebra
(\ref{gtyba}).

Our next example is the well known family \cite{Kulish85} of graded
rational $R$-matrices
\begin{equation} \label{rrm}
     R^{\a \be}_{\g \de} (u,v) = a(u,v) (-1)^{p(\a) p(\be)} \,
				    \de^\a_\g \de^\be_\de +
				    d(u,v) \, \de^\a_\de \de^\be_\g \qd,
\end{equation}
where
\begin{equation} \label{defad}
     a(u,v) = \frac{u - v}{u - v + \i} \qd, \qd
     d(u,v) = \frac{\i}{u - v + \i} \qd.
\end{equation}
$R(u,v)$ solves the Yang-Baxter equation (\ref{ybe}) for arbitrary
matrix dimension $N^2 \times N^2$ and arbitrary grading
$p: \{1, \dots, N\} \rightarrow \mathbb{Z}_2$.

\begin{remark}
Note the following subtlety. The grading introduced in (\ref{rrm}) is
{\em independent} of the grading that enters the definition
(\ref{defej}) of the matrices ${e_j}_\a^\be$. Let $q: \{1, \dots, N\}
\rightarrow \mathbb{Z}_2$ arbitrary. Then, because of the Kronecker
deltas in (\ref{rrm}),
\begin{equation}
     R_{\g \de}^{\a \be} (u,v) = (-1)^{q(\a) + q(\be) + q(\g) + q(\de)}
        \, R_{\g \de}^{\a \be} (u,v) \qd,
\end{equation}
i.e.\ the compatibility condition (\ref{comp}) is satisfied for
arbitrary $p$ and $q$. For example, let $N = 2$, $p(1) = p(2) = 0$. Then
$R(u,v)$ is the $R$-matrix of the XXX spin-$\frac{1}{2}$ Heisenberg
chain, which is compatible with the grading $q(1) = 0$, $q(2) = 1$
leading to a special case of the small polaron Hamiltonian introduced
above.
\end{remark}

Let us now elaborate on the case $p = q$. Since $a(v,v) = 0$ and
$d(v,v) = 1$, the $R$-matrix defined in equation (\ref{rrm}) is
regular. Furthermore,
\begin{equation} \label{hperm}
     \6_u \left. \check R^{\a \be}_{\g \de} (u,v) \right|_{u = v}
	= \i \left[ \de^\a_\g \de^\be_\de - 
	    (-1)^{p(\a) p(\be)} \, \de^\a_\de \de^\be_\g \right] \qd.
\end{equation}
Thus, using $p = q$ in equations (\ref{ham}) and (\ref{hdens}) we find
the Hamiltonian
\begin{equation}
     \hat H = - \sum_{j=1}^L (P_{j j+1} - 1) \qd,
\end{equation}
where $P_{j j+1}$ is the graded permutation operator defined in
(\ref{defp}). Clearly $\hat H$ commutes with the generators
\begin{equation}
     E_\a^\be = \sum_{j=1}^L {e_j}_\a^\be
\end{equation}
of the graded Lie algebra gl($m|n$).

The family of Hamiltonians (\ref{hperm}) based on graded permutations
includes a number of models that are interesting for applications in
physics. In the `non-graded' case ($p(\a) = 0$, $\a = 1, \dots, N$) we
have the XXX spin-$\frac{1}{2}$ chain and its su($N$) generalizations.
The case $m = 1$, $n = 2$ leads us to the supersymmetric $t$-$J$ model
\cite{EsKo92}. In order to see this we shall employ the fermionization
scheme of the previous section. We start with the set of projectors
obtained from the matrix $X_j$ in (\ref{xj}) by deleting its fourth row
and column. We obtain the reduced matrix
\begin{equation} \label{xjtj}
     X_j = \begin{pmatrix}
	     (1 - n_{j \ab})(1 - n_{j \auf}) &
	     (1 - n_{j \ab}) c_{j \auf} &
	     c_{j \ab} (1 - n_{j \auf}) \\
	     (1 - n_{j \ab}) c_{j \auf}^\dagger &
	     (1 - n_{j \ab}) n_{j \auf} &
	     - c_{j \ab} c_{j \auf}^\dagger \\
	     c_{j \ab}^\dagger (1 - n_{j \auf}) &
	     c_{j \ab}^\dagger c_{j \auf} &
	     n_{j \ab} (1 - n_{j \auf})
           \end{pmatrix} \qd.
\end{equation}
The entries $(X_j)^\a_\be$ of this matrix form a complete set of
projection operators on the local space of states spanned by the
basis states $|0\>$, $c_{j \auf}^+ |0\>$, $c_{j \ab}^+ |0\>$. Double
occupancy is excluded on this space of states. Let ${X_j}_\a^\be =
(X_j)^\a_\be$. The operator
\begin{equation}
     {X_j}_\a^\a = 1 - n_{j \auf} n_{j \ab}
\end{equation}
projects the local Hilbert space of electrons onto the space with no
double occupancy. The global projection operator for a chain of $L$
sites is given by the product
\begin{equation} \label{notwo}
     \D = \prod_{j=1}^L (1 - n_{j \auf} n_{j \ab}) \qd.
\end{equation}
The permutation operator $P_{jk}$ is given by equation (\ref{defp})
with ${X_j}_\a^\be$ replacing ${e_j}_\a^\be$. The summation in
(\ref{defp}) is now over three values, $\a, \be = 1, 2, 3$, and the
grading is $p(1) = 0$, $p(2) = p(3) = 1$. An elegant way of taking
into account the simplifications arising from the restriction to the
Hilbert space with with no double occupancy is to consider $P_{jk}
\D$ instead of $P_{jk}$. Since $n_{j \auf} n_{j \ab} \D = 0$, we obtain
\begin{equation} \label{ptj}
     (P_{jk} - 1)\D = \D (c_{j \s}^\dagger c_{k \s} +
        c_{k \s}^\dagger c_{j \s}) \D -
	2 (S_j^a S_k^a - \tst{\4} n_j n_k) \D - (n_j + n_k) \D \qd.
\end{equation}
Here we have introduced the electron density $n_j = n_{j \auf} +
n_{j \ab}$ and the spin densities
\begin{equation} \label{defs}
     S_j^a = \tst{\2} \s_{\a \be}^a c_{j \a}^\dagger c_{j \be} \qd.
\end{equation}
The $\s^a$, $a = x, y, z$, are the Pauli matrices, and we identify
1 with $\auf$ an 2 with $\ab$ in the summation over $\a$ and $\be$.
Inserting (\ref{ptj}) into the expression (\ref{hperm}) for the
Hamiltonian we obtain the familiar Hamiltonian of the supersymmetric
$t$-$J$ model \cite{BrRi70,ZhRi88,Schlottmann87,BaBl90,Sarkar91,%
EsKo92,FoKa93}.

Let us also write down the corresponding $L$-matrix, which follows
from equation (\ref{defgl}),
\begin{equation}
     \CL_j (u,v) = a(u,v) + d(u,v) \begin{pmatrix}
		    {X_j}_1^1 & {X_j}_2^1 & {X_j}_3^1 \\
		    {X_j}_1^2 & - {X_j}_2^2 & - {X_j}_3^2 \\
		    {X_j}_1^3 & - {X_j}_2^3 & - {X_j}_3^3
                 \end{pmatrix} \qd.
\end{equation}
This form of the $L$-matrix suggests a similar form for the monodromy
matrix of the corresponding inhomogeneous model,
\begin{equation} \label{tjmono}
     \CT (u; \xi_1, \dots, \xi_L) = \begin{pmatrix}
	  A(u) & B_1 (u) & B_2 (u) \\
	  C_1 (u) & D^1_1 (u) & D^1_2 (u) \\
	  C_2 (u) & D^2_1 (u) & D^2_2 (u)
        \end{pmatrix} \qd.
\end{equation}
\section{Solution of the quantum inverse problem}
We are now ready to formulate our main result, which is a formula that
expresses the local projection matrices ${e_j}_\a^\be$ for fundamental
graded models in terms of the elements of the monodromy matrix. We shall
assume we are given a solution of the Yang-Baxter equation (\ref{ybe})
which is regular and unitary. Unitarity means that $R(u,v)$ satisfies
the equation
\begin{equation} \label{unitary}
     R^{\a \be}_{\g \de} (u,v) R^{\de \g}_{\a' \be'} (v,u) =
	\de^\a_{\be'} \de^\be_{\a'} \qd.
\end{equation}
Let $p$ be a grading that is compatible with the $R$-matrix in the
sense of equation (\ref{comp}), and let $\CT(u) = \CT(u; \xi_1, \dots,
\xi_L)$ be the corresponding inhomogeneous monodromy matrix
(\ref{tin}). Then we have the following formula,
\begin{equation} \label{invers}
     {e_n}_\a^\be = (-1)^{p(\a) p(\be)} \prod_{j=1}^{n-1}
		    \str(\CT(\x_j)) \cdot \CT_\a^\be (\x_n) \cdot
		    \prod_{j = n+1}^L \str(\CT(\x_j)) \qd.
\end{equation}
Equation (\ref{invers}) is our main result. It constitutes a solution
of the quantum inverse problem for fundamental graded models. We shall
prove it in the remaining sections of this article. For $m = 2$, $n = 0$
($p(1) = p(2) = 0$) equation (\ref{invers}) reduces to a result recently
obtained by Kitanine et al.\ \cite{KMT99a}. Note that because of
(\ref{trans}) no ordering is required for the products on the right hand
side of (\ref{invers}).

Before proceeding with the proof of (\ref{invers}) let us illustrate
the equation through the examples of the previous section. Note that the
functions $b(u,v)$, $c(u,v)$ in (\ref{defbc}) and $a(u,v)$, $d(u,v)$ in
(\ref{defad}) have been chosen in such a way that the corresponding
$R$-matrices (\ref{rxxz}), (\ref{rrm}) satisfy the unitarity condition~%
(\ref{unitary}).

For the small polaron model the monodromy matrix is of the form
\begin{equation}
     \CT(u) = \begin{pmatrix} A(u) & B(u) \\ C(u) & D(u) \end{pmatrix}
	      \qd.
\end{equation}
Using the fermionization (\ref{xj1}) we obtain from (\ref{invers}),
\begin{align} \label{spinvers1}
     c_n^+ = & \prod_{j=1}^{n-1} \bigl( A(\x_j) - D(\x_j) \bigr)
		  \cdot B (\x_n) \cdot
		  \prod_{j = n+1}^L \bigl( A(\x_j) - D(\x_j) \bigr)
		  \qd, \\
     c_n = & \prod_{j=1}^{n-1} \bigl( A(\x_j) - D(\x_j) \bigr)
		  \cdot C (\x_n) \cdot
		  \prod_{j = n+1}^L \bigl( A(\x_j) - D(\x_j) \bigr)
		  \qd, \\
     n_n = - & \prod_{j=1}^{n-1} \bigl( A(\x_j) - D(\x_j) \bigr)
		  \cdot D (\x_n) \cdot
		  \prod_{j = n+1}^L \bigl( A(\x_j) - D(\x_j) \bigr) \qd.
		  \label{spinvers3}
\end{align}

A similar set of equations holds for the local operators of the
supersymmetric $t$-$J$ model. The monodromy matrix was presented in
equation (\ref{tjmono}). Since the grading is $p(1) = 0$, $p(2) = p(3)
= 1$, we have $\str(\CT(u)) = A(u) - \tr(D(u))$, and
\begin{align} \label{tj1}
     (1 - n_{n \ab}) c_{n \auf}^+ =
	& \prod_{j=1}^{n-1} \str(\CT(\x_j)) \cdot B_1 (\x_n)
		  \cdot \prod_{j = n+1}^L \str(\CT(\x_j)) \qd, \\
		     \label{tj2}
     c_{n \ab}^+ (1 - n_{n \auf}) =
	& \prod_{j=1}^{n-1} \str(\CT(\x_j)) \cdot B_2 (\x_n)
		  \cdot \prod_{j = n+1}^L \str(\CT(\x_j)) \qd, \\
		     \label{tj3}
     (1 - n_{n \ab}) c_{n \auf} =
	& \prod_{j=1}^{n-1} \str(\CT(\x_j)) \cdot C_1 (\x_n)
		  \cdot \prod_{j = n+1}^L \str(\CT(\x_j)) \qd, \\
		     \label{tj4}
     c_{n \ab} (1 - n_{n \auf}) =
	& \prod_{j=1}^{n-1} \str(\CT(\x_j)) \cdot C_2 (\x_n)
		  \cdot \prod_{j = n+1}^L \str(\CT(\x_j)) \qd,\\
		     \label{tj5}
     S_n^+ =
	- & \prod_{j=1}^{n-1} \str(\CT(\x_j)) \cdot D^2_1 (\x_n)
		  \cdot \prod_{j = n+1}^L \str(\CT(\x_j)) \qd, \\
		     \label{tj6}
     S_n^- =
	- & \prod_{j=1}^{n-1} \str(\CT(\x_j)) \cdot D^1_2 (\x_n)
		  \cdot \prod_{j = n+1}^L \str(\CT(\x_j)) \qd, \\
		     \label{tj7}
     S_n^z =
	- \frac{1}{2} & \prod_{j=1}^{n-1} \str(\CT(\x_j)) \cdot
	      \bigl( D^1_1 (\x_n) - D^2_2 (\x_n) \bigr)
		  \cdot \prod_{j = n+1}^L \str(\CT(\x_j)) \qd, \\
		     \label{tj8}
     (1 - n_{n \ab}) (1 - n_{n \auf}) =
	& \prod_{j=1}^{n-1} \str(\CT(\x_j)) \cdot A (\x_n)
		  \cdot \prod_{j = n+1}^L \str(\CT(\x_j)) \qd.
\end{align}
On the restricted Hilbert space of the supersymmetric $t$-$J$ model,
where double occupancy of lattice sites is excluded, the operators on
the left hand side of equations (\ref{tj1})-(\ref{tj4}) are appropriate
creation and annihilation operators. The operator $(1 - n_{n \ab})
c_{n \auf}^+$, for instance, creates an up-spin electron on site $n$,
provided this site is not occupied by a down-spin electron. The local
spin operators $S_n^a$ ($a = x, y, z$) were introduced in equation
(\ref{defs}). $S_n^+$ and $S_n^-$ in (\ref{tj5}) and (\ref{tj6}) are
defined as $S_n^+ = S_n^x + \i S_n^y = c_{n \auf}^+ c_{n \ab}$ and
$S_n^- = S_n^x - \i S_n^y = c_{n \ab}^+ c_{n \auf}$. These operators
induce a local spin flip. The operator $(1 - n_{n \ab})(1 - n_{n \auf})$
on the left hand side of (\ref{tj8}) acts as $1 - (n_{n \auf} +
n_{n \ab})$ on the restricted Hilbert space of the supersymmetric
$t$-$J$ model and thus essentially gives the local particle number
operator.

Note that equations (\ref{tj1})-(\ref{tj8}) after appropriate
replacement of the monodromy matrix also apply to the `$t$-$0$ model'
(the infinite coupling limit of the Hubbard model below half-filling)
which was solved by algebraic Bethe ansatz in \cite{GoMu98}.

We would like to stress, that our main result, equation (\ref{invers}),
also holds for homogeneous, translationally invariant models, for
which $\x_j = v = 0$, for $j = 1, \dots, L$. In this case (\ref{invers})
takes a particularly simple form, since $\str (\CT(0)) = \hat U$,
where $\hat U$ is the homogeneous shift operator (\ref{shift}). Using
this fact equation (\ref{invers}) turns into
\begin{equation}
     {e_n}_\a^\be = (-1)^{p(\a) p(\be)} \,
		       \hat U^{n-1} \CT_\a^\be (0) \, \hat U^{L-n} \qd.
\end{equation}
Performing a similarity transformation with $\hat U^{1-n}$ we obtain
the amazingly simple result
\begin{equation}
     {e_1}_\a^\be = (-1)^{p(\a) p(\be)} \,
		       \CT_\a^\be (0) \, \hat U^{-1} \qd,
\end{equation}
which in section 1 was obtained for the special case of the XYZ chain.
To give another example, equations (\ref{spinvers1})-(\ref{spinvers3}),
for instance, are in the homogeneous case equivalent to
\begin{equation}
     c_1^+ = B(0) \, \hat U^{-1} \qd, \qd c_1 = C(0) \, \hat U^{-1}
	\qd, \qd n_1 = - D(0) \, \hat U^{-1} \qd.
\end{equation}
\section{\boldmath The fermionic $R$-operator}
The role of the matrix $\check R(u,v)$ in the graded Yang-Baxter algebra
(\ref{gyba}) is to switch the order of the two auxiliary spaces. The
definition of an operator that similarly switches the order of quantum
spaces in a product of two $L$-matrices requires appropriate use of the
grading. Recently, such an operator was introduced for several important
models by Umeno, Shiroishi and Wadati \cite{USW98a,USW98b} and was
called fermionic $R$-operator. Here we give a general definition of the
fermionic $R$-operator associated with a solution $R(u,v)$ of the
Yang-Baxter equation (\ref{ybe}). For a given grading and a solution
$R(u,v)$ of the Yang-Baxter equation (\ref{ybe}) that is compatible
with this grading (cf (\ref{comp})) we define
\begin{equation} \label{defrf}
     \CR^f_{jk} (u,v) = (-1)^{p(\g) + p(\a)(p(\be) + p(\g))} \,
			   R^{\a \be}_{\g \de} (u,v) {e_j}_\a^\g
			   {e_k}_\be^\de \qd.
\end{equation}
The fermionic $R$-operator will be an important tool in the proof of
our main result (\ref{invers}). Let us summarize its properties in
the following lemma.
\begin{lemma}
Properties of the fermionic $R$-operator.
\begin{enumerate}
\item
Evenness. The fermionic $R$-operator is even,
\begin{equation} \label{rfeven}
     p(\CR^f_{jk} (u,v)) = 0 \qd.
\end{equation}
\item
Bilinear relation. The fermionic $R$-operator satisfies
\begin{equation} \label{rfbe}
     \CR^f_{jk} (\x_j, \x_k) \CL_k (u, \x_k) \CL_j (u, \x_j) =
        \CL_j (u, \x_j) \CL_k (u, \x_k) \CR^f_{jk} (\x_j, \x_k) \qd.
\end{equation}
\item
Yang-Baxter equation. The fermionic $R$-operator satisfies the
following form of the Yang-Baxter equation,
\begin{equation} \label{rfybe}
     \CR^f_{12} (u,v) \CR^f_{13} (u,w) \CR^f_{23} (v,w) =
        \CR^f_{23} (v,w) \CR^f_{13}(u,w) \CR^f_{12}(u,v) \qd.
\end{equation}
\item
Regularity. If $R(u,v)$ is regular, say $R^{\a \be}_{\g \de} (v,v) =
\de^\a_\de \de^\be_\g$, then
\begin{equation} \label{rfreg}
     \CR^f_{jk} (v,v) = P_{jk} \qd,
\end{equation}
where $P_{jk}$ is the graded permutation operator (\ref{defp}).
\item
Unitarity. If $R(u,v)$ is unitary (cf equation (\ref{unitary})), then
$\CR^f_{jk} (u,v)$ is unitary in the sense that
\begin{equation} \label{rfuni}
     \CR^f_{jk} (u,v) \CR^f_{kj} (v,u) = \id \qd.
\end{equation}
\end{enumerate}
\end{lemma}
\begin{proof}
\begin{enumerate}
\item
The evenness of the fermionic $R$-operator is a direct consequence of
the compatibility condition (\ref{comp}).
\item
Using the commutation relations (\ref{coantico}) and the projection
property (\ref{samesite}) for the matrices ${e_j}_\a^\be$ as well as 
the compatibility condition (\ref{comp}) and the Yang-Baxter equation
(\ref{ybe}), the matrix elements on both sides of (\ref{rfbe}) can be
reduced to
\[
\begin{split}
     & (-1)^{\{ p(\a) p(\be) + p(\be) p(\g) + p(\g) p(\a) +
	      p(\be'') (p(\g) + p(\g'')) \}} \\ & \qqqd
		R_{\be' \g'}^{\be \g} (\x_1,\x_2)
		R_{\a' \g''}^{\a \g'} (u,\x_2)
		R_{\a'' \be''}^{\a' \be'} (u,\x_1)
		{e_1}_\be^{\be''} {e_2}_\g^{\g''} \qd.
\end{split}
\]
\item
The proof is similar to the proof of (ii). Using (\ref{coantico}),
(\ref{samesite}) and (\ref{comp}), (\ref{ybe}) both sides of equation
(\ref{rfybe}) reduce to
\[
\begin{split}
     & (-1)^{\{ p(\a) p(\be) + p(\be) p(\g) + p(\g) p(\a) +
	      p(\a'') (p(\a) + p(\a'')) + p(\be'') (p(\g) + p(\g'')) \}}
		\\ & \qqqd
		R_{\be' \g'}^{\be \g} (\x_1,\x_2)
		R_{\a' \g''}^{\a \g'} (u,\x_2)
		R_{\a'' \be''}^{\a' \be'} (u,\x_1)
		{e_1}_\a^{\a''} {e_2}_\be^{\be''} {e_3}_\g^{\g''} \qd.
\end{split}
\]
\item
\[
     (-1)^{p(\g) + p(\a) (p(\be) + p(\g))} \de^\a_\de \de^\be_\g
		{e_j}_\a^\g {e_k}_\be^\de = (-1)^{p(\be)} 
		{e_j}_\a^\be {e_k}_\be^\a = P_{jk} \qd.
\]
\item
Using (\ref{coantico}), (\ref{samesite}) and (\ref{comp}) we obtain
\[
     \CR^f_{jk} (u,v) \CR^f_{kj} (v,u) = (-1)^{p(\be)(p(\a) + p(\be'))}
	R^{\a \be}_{\g \de} (u,v) R^{\de \g}_{\a' \be'} (v,u)
		{e_j}_\a^{\be'} {e_k}_\be^{\a'} \qd,
\]
and the assertion follows from (\ref{unitary}).
\end{enumerate}
\end{proof}
\section{The shift operator}
In this section we shall use the fermionic $R$-operator introduced
above in order to define the shift operator for inhomogeneous graded
models. We shall explore the properties of the shift operator and
shall eventually use these properties to prove our main result
(\ref{invers}).

Let us start with a slightly more general concept. The inhomogeneous
monodromy matrix defined in (\ref{tin}) is an {\em ordered} product
of $L$-matrices. In the following we shall indicate the order of the
factors by supplying subscripts to the monodromy matrix
\begin{equation} \label{tin2}
     \CT_{1 \dots L} (u; \xi_1, \dots, \xi_L) =
	\CT(u; \xi_1, \dots, \xi_L) = \CL_L (u, \xi_L)
				      \dots \CL_1 (u, \xi_1) \qd.
\end{equation}
As can be seen from (\ref{rfeven}) and (\ref{rfbe}) the fermionic
$R$-operator $\CR^f_{j j+1} (\x_j,\x_{j+1})$ interchanges the two
neighbouring factors $\CL_{j+1} (u,\x_{j+1})$ and $\CL_j (u,\x_j)$ in
the monodromy matrix. Since the symmetric group $\mathfrak{S}^L$ is
generated by the transpositions of nearest neighbours, the $L$-matrices
on the right hand side of (\ref{tin2}) can be arbitrarily reordered by
application of an appropriate product of fermionic $R$-operators. This
means that for every $\tau \in \mathfrak{S}^L$ there exists an operator
$\CR^\tau_{1 \dots L} (\x_1, \dots, \x_L)$, which is a product of
fermionic $R$-operators and induces the action of the permutation
$\tau \in \mathfrak{S}^L$ on the inhomogeneous monodromy matrix,
\begin{multline} \label{rtau}
     \CR^\tau_{1 \dots L} (\x_1, \dots, \x_L)
	\CT_{1 \dots L} (u; \x_1, \dots, \x_L) = \\
	\CT_{\tau (1) \dots \tau (L)}
	     (u; \x_{\tau (1)}, \dots, \x_{\tau (L)})
        \CR^\tau_{1 \dots L} (\x_1, \dots, \x_L) \qd.
\end{multline}
The non-graded analogue of this operator was introduced in
\cite{MaSa96}.

Let us construct $\CR^\tau_{1 \dots L} (\x_1, \dots, \x_L)$ explicitly.
We shall use the shorthand notation $\CR^\tau_{1 \dots L} =
\CR^\tau_{1 \dots L} (\x_1, \dots, \x_L)$, $\CT_{1 \dots L} (u) =
\CT_{1 \dots L} (u; \x_1, \dots, \x_L)$ and $\CR^f_{j k} = \CR^f_{j k}
(\x_j, \x_k)$ whenever the order of the inhomogeneities $\x_1, \dots,
\x_L$ is the same as the order of the corresponding lattice sites. For
$j = 1, \dots, L - 1$ define $\s_j \in \mathfrak{S}^L$ by
\begin{equation}
     \s_j (k) = \begin{cases}
		   j + 1 & \text{if $k = j$,} \\
		   j & \text{if $k = j + 1$,} \\
		   k & \text{else.}
	        \end{cases}
\end{equation}
The $\s_j \in \mathfrak{S}^L$ are transpositions of nearest neighbours.
It follows from (\ref{rfeven}), (\ref{rfbe}) that
\begin{equation}
     \CR^f_{j j+1} \, \CT_{1 \dots L} (u) =
        \CT_{\s_j (1) \dots \s_j (L)} (u) \, \CR^f_{j j+1}\qd.
\end{equation}
This means that $\CR^{\s_j}_{1 \dots L} = \CR^f_{j j+1}$. Choose
$\tau \in \mathfrak{S}^L$ arbitrarily. Then
\begin{equation} \label{ttrans}
     \CR^f_{\tau (j), \tau (j+1)} \,
	\CT_{\tau (1) \, \dots \, \tau (L)} (u) =
	\CT_{\tau \s_j (1) \, \dots \, \tau \s_j (L)} (u) \,
        \CR^f_{\tau (j), \tau (j+1)} \qd.
\end{equation}
Since the transpositions of nearest neighbours $\s_j$, $j = 1, \dots,
L - 1$, generate the symmetric group $\mathfrak{S}^L$, there is a
finite sequence $(j_p)_{p=1}^n$, such that $\tau = \s_{j_1} \dots
\s_{j_n}$. Let $\tau_p = \s_{j_1} \dots \s_{j_p}$, $p = 1, \dots, n$
and $\tau_0 = \id$. Then $\tau = \tau_n$, and, using (\ref{ttrans}), we
conclude that
\begin{equation} \label{ttrans2}
     \CR^f_{\tau_{p-1} (j_p), \tau_{p-1} (j_p + 1)} \,
	\CT_{\tau_{p-1} (1) \, \dots \, \tau_{p-1} (L)} (u) =
	\CT_{\tau_p (1) \, \dots \, \tau_p (L)} (u) \,
        \CR^f_{\tau_{p-1} (j_p), \tau_{p-1} (j_p + 1)} \qd,
\end{equation}
for $p = 1, \dots, n$. By iteration of the latter equation we obtain
\begin{multline} \label{tperm}
     \CR^f_{\tau_{n-1} (j_n), \tau_{n-1} (j_n + 1)} \dots
        \CR^f_{\tau_1 (j_2), \tau_1 (j_2 + 1)} \CR^f_{j_1, j_1 + 1} \,
	\CT_{1 \dots L} (u) = \\
	\CT_{\tau (1) \, \dots \, \tau (L)} (u) \,
        \CR^f_{\tau_{n-1} (j_n), \tau_{n-1} (j_n + 1)} \dots
        \CR^f_{\tau_1 (j_2), \tau_1 (j_2 + 1)} \CR^f_{j_1, j_1 + 1} \qd.
\end{multline}
Thus we have constructed an explicit expression for the operator
$\CR^\tau_{1 \dots L}$,
\begin{equation} \label{rtauex}
     \CR^\tau_{1 \dots L} =
	  \CR^f_{\tau_{n-1} (j_n), \tau_{n-1} (j_n + 1)} \dots
	  \CR^f_{\tau_1 (j_2), \tau_1 (j_2 + 1)} \CR^f_{j_1, j_1 + 1}
	     \qd.
\end{equation}

Let us now specify to the case, when $\tau$ is equal to the cyclic
permutation $\g = \s_1 \dots \s_{L-1}$. Then $j_p = p$, $p = 1, \dots,
L- 1$, in our above construction, and $\g_p = \s_1 \dots \s_p$. Thus
$\g_{p-1} (j_p) = \g_{p-1} (p) = 1$, and $\g_{p-1} (j_p + 1) =
\g_{p-1} (p + 1) = p + 1$. Using (\ref{rtauex}) we obtain
\begin{equation}
     \CR^\g_{1 \dots L} = \CR^f_{1L} \CR^f_{1 L-1} \dots \CR^f_{12} \qd.
\end{equation}
The operator $\CR^\g_{1 \dots L}$ induces a shift by one site on the
inhomogeneous mono\-dromy matrix. Now (\ref{ttrans}) implies that
\begin{equation}
     \CR^\g_{\g (1) \dots \g (L)} \CT_{\g (1) \dots \g (L)} (u) =
	\CT_{\g^2 (1) \dots \g^2 (L)} (u) \CR^\g_{\g (1) \dots \g (L)}
	   \qd.
\end{equation}
It follows by multiplication by $\CR^\g_{1 \dots L}$ from the right,
that
\begin{equation}
     \CR^{\g^2}_{1 \dots L} = \CR^\g_{\g (1) \dots \g (L)}
	\CR^\g_{1 \dots L} \qd.
\end{equation}
Iterating the above steps we arrive at the following lemma.
\begin{lemma}
The operator
\begin{equation} \label{shiftin}
     \CR^{\g^n}_{1 \dots L} =
	\CR^\g_{\g^{n-1} (1) \dots \g^{n-1} (L)}
	\CR^\g_{\g^{n-2} (1) \dots \g^{n-2} (L)} \dots
	\CR^\g_{1 \dots L} \qd,
\end{equation}
where
\begin{equation} \label{rgg}
     \CR^\g_{\g^{p-1} (1) \dots \g^{p-1} (L)} = \CR^f_{p p-1} \dots
	\CR^f_{p1} \CR^f_{pL} \dots \CR^f_{p p+1} \qd,
\end{equation}
generates a shift by $n$ sites on the inhomogeneous lattice, i.e.
\begin{equation} \label{nshift}
     \CR^{\g^n}_{1 \dots L} \, \CT_{1 \dots L} (u) =
     \CT_{n+1 \dots L 1 \dots n} (u) \, \CR^{\g^n}_{1 \dots L} \qd.
\end{equation}
\end{lemma}

Since $\g^L = \id$, we conclude from (\ref{nshift}) that
\begin{equation}
     \CR^{\g^L}_{1 \dots L} \, \CT_{1 \dots L} (u) =
     \CT_{1 \dots L} (u) \, \CR^{\g^L}_{1 \dots L} \qd.
\end{equation}
If $\CR^f_{jk}$ is unitary, we have the following stronger result.
\begin{lemma}
Let $\CR^f_{jk}$ be unitary (cf equation \ref{rfuni}). Then
\begin{equation}
     \CR^{\g^L}_{1 \dots L} = \id \qd.
\end{equation}
\end{lemma}
\begin{proof}
Let us first prove the case $L=2$. Then $\CR^\g_{12} = \CR^f_{12}$ and
$\CR^{\g^2}_{12} = \CR^\g_{\g 1 \g 2} \CR^\g_{12} = \CR^\g_{21}
\CR^\g_{12} = \CR^f_{21} \CR^f_{12} = \id$. The last equation holds,
since by hypothesis, $\CR^f_{12}$ is unitary.

For the case $L > 2$ we start from the Yang-Baxter equation
(\ref{rfybe}),
\begin{equation} \label{ybele3}
     \CR^f_{L, L-n} \CR^f_{L, j} \CR^f_{L-n, j} =
        \CR^f_{L-n, j} \CR^f_{L, j} \CR^f_{L, L-n} \qd.
\end{equation}
By iterated use of (\ref{ybele3}) we obtain
\begin{multline} \label{multrf}
     \CR^f_{L, L-n} (\CR^f_{L, L-n-1} \dots \CR^f_{L, 1})
                    (\CR^f_{L-n, L-n-1} \dots \CR^f_{L-n, 1}) = \\
                    (\CR^f_{L-n, L-n-1} \dots \CR^f_{L-n, 1})
		    (\CR^f_{L, L-n-1} \dots \CR^f_{L, 1})
		       \CR^f_{L, L-n} \qd,
\end{multline}
for $n = 1, \dots, L- 2$.

Let us introduce the truncated cyclic permutations $\g_p = \s_1 \dots
\s_{p-1}$, $p = 2, \dots, L$, as above. $\g_p$ induces a cyclic shift
on the $p$-tuple $(1, \dots, p)$ and leaves the $(L - p)$-tuple
$(p + 1, \dots, L)$ invariant. Using (\ref{multrf}), it follows that
\begin{equation}
\begin{split}
     \CR^f_{L, L-n} & \dots \CR^f_{L, 1} \,
	\CR^\g_{\g^{L-n-1} (1) \, \dots \, \g^{L-n-1} (L)} \\
        & = \CR^f_{L, L-n} \, (\CR^f_{L, L-n-1} \dots \CR^f_{L, 1})
                    (\CR^f_{L-n, L-n-1} \dots \CR^f_{L-n, 1}) \cdot \\
		    & \qqqd \qqqd \qqqd \qqd \cdot
                    (\CR^f_{L-n, L} \dots \CR^f_{L-n, L-n+1}) \\
        & = (\CR^f_{L-n, L-n-1} \dots \CR^f_{L-n, 1})
	       (\CR^f_{L, L-n-1} \dots \CR^f_{L, 1}) \cdot \\
		    & \qqqd \qqqd \qqd \cdot
	       \underbrace{\CR^f_{L, L-n} \CR^f_{L-n, L}}_%
		  {\textstyle = \, \id} \,
                    (\CR^f_{L-n, L-1} \dots \CR^f_{L-n, L-n+1}) \\
        & = (\CR^f_{L-n, L-n-1} \dots \CR^f_{L-n, 1})
               (\CR^f_{L-n, L-1} \dots \CR^f_{L-n, L-n+1}) \cdot \\
		    & \qqqd \qqqd \qqqd \qqqd \cdot
	              (\CR^f_{L, L-n-1} \dots \CR^f_{L, 1}) \\
	& = \CR^{\g_{L-1}}_{\g^{L-n-1}_{L-1} (1), \, \dots \, ,
			    \g^{L-n-1}_{L-1} (L-1), \, L} \,
	    \CR^f_{L, L-n-1} \dots \CR^f_{L, 1} \qd. \raisetag{3.5ex}
\end{split}
\end{equation}
Hence,
\begin{equation}
\begin{split}
   \CR^{\g^L}_{1 \dots L} & = \CR^\g_{\g^{L-1} (1) \dots \g^{L-1} (L)}
			         \,
			      \CR^\g_{\g^{L-2} (1) \dots \g^{L-2} (L)}
			      \dots \CR^\g_{1 \dots L} \\
                          & = \CR^f_{L, L-1} \dots \CR^f_{L, 1} \,
			      \CR^\g_{\g^{L-2} (1) \dots \g^{L-2} (L)}
				 \,
			      \CR^\g_{\g^{L-3} (1) \dots \g^{L-3} (L)}
			      \dots \CR^\g_{1 \dots L} \\
                          & = \CR^{\g_{L-1}}_{\g^{L-2}_{L-1} (1), \,
				   \dots \, , \g^{L-2}_{L-1} (L-1),
				   \, L} \,
			      \CR^f_{L, L-2} \dots \CR^f_{L, 1} \,
			      \CR^\g_{\g^{L-3} (1) \dots \g^{L-3} (L)}
			      \dots \CR^\g_{1 \dots L} \\
                          & = \CR^{\g^{L-1}_{L-1}}_{1 \dots L} =
                              \CR^{\g^{L-2}_{L-2}}_{1 \dots L} = \dots
			      = \CR^{\g^{2}_{2}}_{1 \dots L} \qd.
\end{split}
\end{equation}
Since $\g_2 = \s_1$ and $\CR^{\s_1}_{1 \dots L} = \CR^f_{12}$, the
latter equation reduces the proof of Lemma 3 for $L > 2$ to the case
$L = 2$, which was proved above.
\end{proof}
Our next lemma can be used to establishes a connection between the
inhomogeneous monodromy matrix (\ref{tin2}) and the shift operator
(\ref{shiftin})
\begin{lemma}
Let $X = X^\a_\be e_\a^\be \in \End (\mathbb{C}^{m+n})$ and let
$R(u,v)$ be regular, say, $R^{\a \be}_{\g \de} (v,v) = \de^\a_\de
\de^\be_\g$. Then
\begin{equation} \label{strxt}
     \str (X \CT_{n \dots L 1 \dots n-1} (\x_n)) =
	(-1)^{p(\a) + p(\a) p(\be)} X^\a_\be {e_n}_\a^\be \,
	\CR^\g_{\g^{n-1} 1 \dots \g^{n-1} L} \qd.
\end{equation}
\end{lemma}
\begin{proof}
\begin{align}
    \str( X & \CT_{n \dots L 1 \dots n-1} (\x_n) ) \notag \\
      & = (-1)^{p(\a)} X^\a_\be \,
	     {\CL_{n-1}}^\be_{\be_{n-1}} (\x_n, \x_{n-1}) \dots
	     {\CL_1}^{\be_2}_{\be_1} (\x_n, \x_1)
	     {\CL_L}^{\be_1}_{\be_L} (\x_n, \x_L) \cdot \notag \\
      & \qqqd \qqqd \qqd
	\dots {\CL_{n+1}}^{\be_{n+2}}_{\be_{n+1}} (\x_n, \x_{n+1}) \,
	(-1)^{p(\a) p(\be_{n+1})} \, {e_n}_\a^{\be_{n+1}} \notag \\
      & = (-1)^{\left\{ p(\a) + p(\a) p(\be) +
		\sum_{\genfrac{}{}{0pt}{2}{j=1}{j \ne n}}^L
		      ( p(\be_j) + p(\a_j) p(\be_j)) \right\}} \notag \\
      & \qqqd \qqd \cdot X^\a_\be \,
	\de^\be_{\a_{n-1}} \de^{\be_{n-1}}_{\a_{n-2}} \dots
	\de^{\be_2}_{\a_1} \de^{\be_1}_{\a_L} \de^{\be_L}_{\a_{L-1}}
	\dots \de^{\be_{n+2}}_{\a_{n+1}} {e_n}_\a^{\be_{n+1}} \notag \\
      & \qqqd \qqd
	\cdot {\CL_{n+1}}^{\a_{n+1}}_{\be_{n+1}} (\x_n, \x_{n+1}) \dots
	      {\CL_L}^{\a_L}_{\be_L} (\x_n, \x_L)
	      {\CL_1}^{\a_1}_{\be_1} (\x_n, \x_1) \notag \\
	      & \qqqd \qqqd \qqqd \dots
	      {\CL_{n-1}}^{\a_{n-1}}_{\be_{n-1}} (\x_n, \x_{n-1})
		 \notag \\
      & = (-1)^{\left\{ p(\a) + p(\a) p(\be) +
		\sum_{\genfrac{}{}{0pt}{2}{j=1}{j \ne n, n+1}}^L
		      ( p(\be_j) + p(\a_j) p(\be_j)) \right\}} \notag \\
      & \qqqd \qqd \cdot X^\a_\be \, {e_n}_\a^\be
	{e_n}_{\a_{n-1}}^{\be_{n-1}} {e_n}_{\a_{n-2}}^{\be_{n-2}} \dots
	{e_n}_{\a_1}^{\be_1} {e_n}_{\a_L}^{\be_L} \dots
	{e_n}_{\a_{n+2}}^{\be_{n+2}} \notag \\
      & \qqqd \qqd \cdot (-1)^{p(\be_{n+1}) + p(\a_{n+1}) p(\be_{n+1})}
	{e_n}_{\a_{n+1}}^{\be_{n+1}}
	{\CL_{n+1}}^{\a_{n+1}}_{\be_{n+1}} (\x_n, \x_{n+1}) \notag \\
      & \qqqd \qqd
	\cdot {\CL_{n+2}}^{\a_{n+2}}_{\be_{n+2}} (\x_n, \x_{n+2}) \dots
	      {\CL_L}^{\a_L}_{\be_L} (\x_n, \x_L)
	      {\CL_1}^{\a_1}_{\be_1} (\x_n, \x_1) \notag \\
	      & \qqqd \qqqd \qqqd \dots
	      {\CL_{n-1}}^{\a_{n-1}}_{\be_{n-1}} (\x_n, \x_{n-1})
		 \notag \\
      & = (-1)^{\left\{ p(\a) + p(\a) p(\be) +
		\sum_{\genfrac{}{}{0pt}{2}{j=1}{j \ne n, n+1}}^L
		      ( p(\be_j) + p(\a_j) p(\be_j)) \right\}} \notag \\
      & \qqqd \qqd \cdot X^\a_\be \, {e_n}_\a^\be
	{e_n}_{\a_{n-1}}^{\be_{n-1}} {e_n}_{\a_{n-2}}^{\be_{n-2}} \dots
	{e_n}_{\a_1}^{\be_1} {e_n}_{\a_L}^{\be_L} \dots
	{e_n}_{\a_{n+2}}^{\be_{n+2}} \notag \\
      & \qqqd \qqd
	\cdot {\CL_{n+2}}^{\a_{n+2}}_{\be_{n+2}} (\x_n, \x_{n+2}) \dots
	      {\CL_L}^{\a_L}_{\be_L} (\x_n, \x_L)
	      {\CL_1}^{\a_1}_{\be_1} (\x_n, \x_1) \notag \\
	      & \qqqd \qqqd \qqqd \dots
	      {\CL_{n-1}}^{\a_{n-1}}_{\be_{n-1}} (\x_n, \x_{n-1}) \,
	      \CR^f_{n, n+1} \\
      & = (-1)^{p(\a) + p(\a) p(\be)} X^\a_\be {e_n}_\a^\be \,
	  \CR^f_{n, n-1} \dots \CR^f_{n, 1} \CR^f_{n, L} \dots
	     \CR^f_{n, n+1} \notag \\
      & = (-1)^{p(\a) + p(\a) p(\be)} X^\a_\be \, {e_n}_\a^\be
	  \CR^\g_{\g^{n-1} 1 \dots \g^{n-1} L} \qd.
\end{align}
Here we used the regularity in the first equation. In the second
equation we reversed the order of factors and introduced a product
of Kronecker deltas. In the third equation we used the identity
\begin{multline}
     \de^\be_{\a_{n-1}} \de^{\be_{n-1}}_{\a_{n-2}} \dots
	\de^{\be_2}_{\a_1} \de^{\be_1}_{\a_L} \de^{\be_L}_{\a_{L-1}}
	\dots \de^{\be_{n+2}}_{\a_{n+1}} {e_n}_\a^{\be_{n+1}} \\
	   = {e_n}_\a^\be {e_n}_{\a_{n-1}}^{\be_{n-1}}
	     {e_n}_{\a_{n-2}}^{\be_{n-2}} \dots {e_n}_{\a_1}^{\be_1}
	     {e_n}_{\a_L}^{\be_L} \dots {e_n}_{\a_{n+1}}^{\be_{n+1}}
		\qd,
\end{multline}
which follows from (\ref{samesite}). In the fourth equation we used
that
\begin{equation}
     \CR^f_{jk} = (-1)^{p(\be) + p(\a) p(\be)} {e_j}_\a^\be
		  {\CL_k}^\a_\be (\x_j, \x_k)
\end{equation}
and the fact that $\CR^f_{jk}$ is even. In the fifth equation we
iterated the two previous steps of our calculation. Finally in the
sixth equation the formula (\ref{rgg}) entered.
\end{proof}

Setting $X^\a_\be = \de^\a_\be$ in (\ref{strxt}) and using the cyclic
invariance of the super trace we obtain the following corollary to
Lemma 4.
\begin{corollary*}
\begin{equation} \label{strtin}
     \CR^\g_{\g^{n-1} 1 \dots \g^{n-1} L} =
	\str (\CT_{1 \dots L} (\x_n)) \qd.
\end{equation}
\end{corollary*}
Equation (\ref{strtin}) is the inhomogeneous analogue of equation
(\ref{shift}).
\begin{lemma}
We have the following expression for the shift operator in terms of
the elements of the monodromy matrix,
\begin{equation} \label{shiftmono}
     \CR^{\g^n}_{1 \dots L} =
	\prod_{j=1}^n \str (\CT_{1 \dots L} (\x_j)) \qd.
\end{equation}
If $R(u,v)$ is unitary (cf equation (\ref{unitary})), then
$\CR^{\g^n}_{1 \dots L}$ is invertible with inverse
\begin{equation} \label{shiftinv}
     \left( \CR^{\g^n}_{1 \dots L} \right)^{- 1} =
	\prod_{j=n+1}^L \str (\CT_{1 \dots L} (\x_j)) \qd.
\end{equation}
\end{lemma}
\begin{proof}
The lemma follows from Lemma 2, Lemma 3 and Corrolary 1 to Lemma 4.
\end{proof}

We are now prepared to prove our main result, equation (\ref{invers}).
\begin{proof}[Proof of equation (\ref{invers})]
Using Lemma 2, Lemma 4, the corollary to Lemma 4 and Lemma 5 we obtain
\begin{equation}
\begin{split}
     \str (X & \CT_{n \dots L 1 \dots n-1} (\x_n)) \\
	& = \CR^{\g^{n-1}}_{1 \dots L}
	    \str (X \CT_{1 \dots L} (\x_n))
	    \left( \CR^{\g^{n-1}}_{1 \dots L} \right)^{- 1} \\
	    & = \prod_{j=1}^{n-1} \str (\CT_{1 \dots L} (\x_j)) \,
		\cdot \str (X \CT_{1 \dots L} (\x_n)) \cdot
		\prod_{j=n}^L \str (\CT_{1 \dots L} (\x_j)) \\
	    & = (-1)^{p(\a) + p(\a) p(\be)} X^\a_\be {e_n}_\a^\be \,
		\str (\CT_{1 \dots L} (\x_n)) \qd.
\end{split}
\end{equation}
It follows that
\begin{multline}
     (-1)^{p(\a') + p(\a') p(\be')} X^{\a'}_{\be'} {e_n}_{\a'}^{\be'} =
	\\ \prod_{j=1}^{n-1} \str (\CT_{1 \dots L} (\x_j)) \,
	   \cdot \str (X \CT_{1 \dots L} (\x_n)) \cdot
	   \prod_{j=n+1}^L \str (\CT_{1 \dots L} (\x_j)) \qd.
\end{multline}
Finally, by specifying $X^{\a'}_{\be'} = (-1)^{p(\a') + p(\a') p(\be')}
\, \de^{\a'}_\a \de^\be_{\be'}$, we arrive at equation~(\ref{invers}).
\end{proof}

\section*{Summary}
In this article we obtained an explicit solution of the quantum inverse
problem for fundamental graded models. Our main result is the general
formula (\ref{invers}). This formula expresses the local projection
operators ${e_n}_\a^\be$, which represent local spins and local fields,
in terms of the elements of the monodromy matrix. The formula and its
proof essentially simplify for translationally invariant models (all
inhomogeneities coincide, $\x_j = \x$). In the translationally
invariant case the proof is based on the representation of the shift
operator as a product of permutation matrices.

We presented explicit formulae for the solution of the quantum inverse
problem for the XYZ quantum spin chain (\ref{inversxyz1})-%
(\ref{inversxyz3}) and for the supersymmetric $t$-$J$ model of
strongly correlated electrons (\ref{tj1})-(\ref{tj8}). For the XYZ
chain the local projection operators coincide with local Pauli matrices
(quantum spin operators).

We are planing to use our results in forthcoming publications in order
to obtain multiple integral representations for correlation functions
of fundamental graded models.

{\bf Acknowledgment.} The authors would like to thank M. Shiroishi
for valuable communications. This work was supported by the Deutsche
For\-schungsgemeinschaft under grant number Go 825/2-2 (F.G.) and by the
National Science Foundation under grant number PHY-9605226 (V.K.).


\end{document}